\documentclass[usenatbib]{mn2e}
\pdfoutput=1
\usepackage{float}
\usepackage{graphicx}
\usepackage{amsmath}
\usepackage{placeins}
\usepackage{threeparttable}
\usepackage{mathptmx}
\usepackage{txfonts}
\usepackage[T1]{fontenc}
\usepackage{ae,aecompl}
\usepackage{pdfpages}
\usepackage{etex}
\usepackage{tablefootnote}
\usepackage{scalerel}
\usepackage{longtable}
\usepackage{hyperref}
\hypersetup{colorlinks, citecolor=blue}
\newcommand\HI{H\protect\scaleto{$I$}{1.2ex}}
\newcommand{\kms}{\mbox{km\,s$^{-1}$}}
\begin{document}
\title[\HI{} 21 cm observation and mass models of FGC 1440]
{ \HI{} 21 cm observation and mass models of the extremely thin galaxy FGC 1440}

\author[Aditya et al.]
       {K. Aditya$^{1}$ \thanks{E-mail: kaditya@students.iisertirupati.ac.in}, 
       Peter Kamphuis$^{2}$,
       Arunima Banerjee$^{1}$ \thanks{E-mail: arunima@iisertirupati.ac.in},
       Sviatoslav Borisov $^{3, 4, 5}$ \newauthor
       Aleksandr Mosenkov $^{6}$,     
       Aleksandra Antipova $^{5}$,
       Dmitry Makarov $^{5}$\\
$^1$  Department of Physics, Indian Institute of Science Education and Research (IISER) Tirupati, Tirupati - 517507, India\\
$^2$  Ruhr Universität Bochum, Astronomisches Institut, Universitätsstrasse 150, D-44801 Bochum, Germany\\
$^3$  Department of Astronomy, University of Geneva, Chemin Pegasi 51, 1290 Versoix, Switzerland\\
$^4$  Sternberg Astronomical Institute, M.V. Lomonosov Moscow State University, Universitetsky prospect 13, Moscow, 119234 Russia\\
$^5$  Special Astrophysical Observatory, Russian Academy of Sciences, Nizhnii Arkhyz, 369167 Russia\\
$^6$  Department of Physics and Astronomy, N283 ESC, Brigham Young University, Provo, UT 84602, USA}
\maketitle

\begin{abstract}

We present observations and models of the kinematics and distribution of neutral hydrogen (\HI) in the
superthin galaxy FGC 1440 with an optical axial ratio $a/b = 20.4$. Using the Giant Meterwave Radio telescope (GMRT),
we imaged the galaxy with a spectral resolution of 1.7 \kms{} and a spatial resolution of $15\farcs9\times13\farcs5$.
We find that FGC 1440 has an asymptotic rotational velocity of 141.8 \kms. 
The structure of the \HI{} disc in FGC 1440 is that of a typical thin disc warped along the line of sight, but we can not rule out the presence of 
a central thick \HI{} disc.
We find that the dark matter halo in FGC 1440 could be modeled by a 
pseudo- isothermal (PIS) profile with $\rm R_{c}/ R_{d} <2$, where $R_{c}$ is the core radius of the PIS halo and $R_{d}$ the
exponential stellar disc scale length. We note that in spite of the unusually large axial ratio of FGC 1440, the ratio of 
the rotational velocity to stellar vertical velocity dispersion, $\frac{V_{Rot}}{\sigma_{z}} \sim 5 - 8$, 
which is comparable to other superthins. 
Interestingly, unlike previously studied superthin galaxies which are outliers in the $log_{10}(j_{*}) - log_{10}(M_{*})$ relation 
for ordinary bulgeless disc galaxies, FGC 1440 is found to comply with the same. 
The values of $j$ for the stars, gas and the baryons in FGC 1440 are consistent with those of normal spiral galaxies with similar mass.

\end{abstract}


\section{INTRODUCTION}
Superthin galaxies are subset of edge-on low-surface brightness $\rm (\mu_{B}(0)>22.7\,mag\,arcsec^{-2})$ \citep{bothun1997low, mcgaugh1996number} 
disc galaxies with an axial ratio $a/b>10$ \citep{karachentsev2003revised}. These late type disc structures observed at high inclination 
\citep{matthews1999extraordinary, dalcanton2000structural, dalcanton1996chain} are some of least evolved disc structure 
in the Universe \citep{vorontsov1974specification, kautsch2009edge, uson2003hi} characterized by ratio of high gas mass to blue luminosity, 
$\frac{M_{\rm \HI{}}}{L_{B}} \approx 1 M_{\odot}\,L^{-1}_{\odot}$ \citep{goad1981spectroscopic, uson2003hi}, and low star formation rates 
$\sim 0.01 - 0.05 \, M_{\odot}yr^{-1}$ \citep{narayanan2021star, wyder2009star}.

The $\rm \HI{}$ distribution offers interesting clues to the physical processes regulating the structure, dynamics, and evolution of galaxies. 
Recent large $\rm \HI$ surveys, THINGS (The \HI{} Nearby Galaxy Survey) \citep{walter2008things}, LITTLE THINGS (Local Irregulars That Trace Luminosity Extremes, 
The \HI{} Nearby Galaxy Survey) \citep{hunter2012little} have focused on mapping the  $\rm \HI$ distribution in the nearby spiral and dwarf galaxies, and explain the 
role of $\rm \HI$ in regulating the stability/instability and star formation in these galaxies \citep{leroy2008star, bigiel2008star}. The high-resolution rotation 
curves derived in these studies have been used for constraining the dark matter mass in these galaxies \citep{de2008high, oh2015high}. The observed rotation curves
in conjugation with the stellar photometry, see for example; SPARC (Spitzer Photometry and Accurate Rotation Curves) \citep{lelli2016sparc}, and also 
\citep{rubin1980rotational, rubin1985rotation,de2001mass, banerjee2017mass, kurapati2018mass} have been used to test fundamental scaling 
relations between the total acceleration and the acceleration due to baryons \citep{li2018fitting, ghari2019dark, li2019constant}, checking
the consistency of $\rm \Lambda$-CDM model \citep{keller2017lambdacdm} and testing alternate theories of gravity \citep{gentile2011things, naik2019constraints, chan2018testing}. 
High resolution $\rm \HI{}$ observations may help in better understanding and characterizing instabilities like warps, bars and spiral arms, etc. in gaseous discs of the 
galaxies \citep{phookun1993ngc}, also for indirect characterization using Tremaine-Weinberg 
method, see for example \citep{banerjee2013slow, patra2019detection}. This may also help in uncovering previously unseen gaseous companions, see, for example, 
detection of satellite galaxy close to NGC 973 in HEROES (HERschel Observations of Edge-on Spirals)  survey \citep{allaert2015herschel}. The observation of $\rm \HI{}$ halos see 
for example HALOGAS (Hydrogen Accretion in LOcal GAlaxieS) survey \citep{heald2011westerbork} in galaxies provides us with important pointers for 
understanding the accretion mechanism for replenishing the gas needed for star formation \citep{voigtlander2013kinematics, zschaechner2012halogas, kamphuis2008structure},  
also see EDGE(Extragalactic Database for Galaxy Evolution survey ) - CALIFA (Calar Alto Legacy Integral Field Area) survey \citep{barrera2021edge, bolatto2017edge}. \\

Measurement of the $\rm \HI{}$ distribution along with the $\rm \HI{}$ dispersion and the radial variation of the scaleheight can be used to model the 
shape of the dark matter \citep{olling1995usage, olling1996highly}. The method was applied to investigate the shape of dark matter halo of eight 
edge-on gas-rich galaxies with $a/b>10$ by \cite{o2010dark, peters2017shape}. Dark matter halo plays an essential role in regulating the structure of 
the stellar disc; previous studies by \cite{banerjee2013some, banerjee2010dark} have shown that a compact dark matter halo plays a vital role in regulating 
the superthin structure and in stabilizing the galactic disc against axis-symmetric instability \citep{garg2017origin, van2001kinematics, ghosh2014suppression}. 
Since these galaxies are rich in neutral hydrogen, it contributes significantly to the total potential and plays an important role in regulating the vertical structure of both the stellar and the $\rm \HI{}$ disc \citep{narayan2002vertical}. Similarly, the role of $\rm \HI{}$ gas on the stability of galaxy disc was investigated by \cite{jog1996local, romeo2013simple}. Recent studies have pointed out that the low value of the observed stellar scaleheight is a direct outcome of 
very low vertical stellar velocity dispersion and that these superthin galaxies are highly stable despite low values of dispersion \citep{10.1093/mnras/stab155}. 

Of all the well known superthin galaxies extensively studied in literature till now for the structure of neutral hydrogen, stellar dynamics and properties of 
dark matter haloes ; UGC 7321 \citep{uson2003hi, matthews2003high, matthews1999extraordinary, banerjee2010dark, sarkar2019flaring, 10.1093/mnras/stab155,
komanduri2020dynamical, pohlen2003evidence} has the highest axial ratio $a/b$ equal to 15.4, followed by FGC 1540 \citep{kurapati2018mass} which has $a/b$ equal to 12.25. 
Other superthin galaxies IC 2233 \citep{matthews2008corrugations,matthews2007h, gallagher1976surface} and IC 5249 
\citep{abe1999observation, van2001kinematics, byun1998surface,  yock1999observation}  have an a/b equal to 8.9 and 10.4 respectively. 

In this paper, we report $\rm \HI{}$ 21 cm observation of FGC 1440, which has
an axial ratio equal to 20.36 ($B$-band) and is among the flattest known galaxies. To our knowledge, no previous studies have mapped the distribution of the neutral hydrogen studied shape of dark matter halo in such extremely thin galaxies; $a/b\geq20$. We derive the basic structural 
and kinematic properties of FGC 1440 by modeling the three-dimensional distribution of the neutral hydrogen in FGC 1440. Comparing different model 
datacubes, we derive limits on the scaleheight, velocity dispersion, and inclination. We finally use the total rotation curve in conjugation with the stellar photometry to derive constraints on the shape of dark matter density in FGC 1440. Further, using the stellar and 
the \HI{} surface density, along with the dark matter mass models, we solve the two-component Jeans equation for modeling the vertical stellar dispersion as a function of radius. We use the observed stellar scaleheight and the limits on the \HI{} velocity dispersion 
and the \HI{} scaleheight as constraints on the two-component Jeans equation \citep{narayan2002vertical, banerjee2010dark, sarkar2020general, sarkar2019vertical,
patra2020theoretical, patra2020h, patra2018molecular}.

The paper is organized as follows; in \S 2, we introduce the target FGC 1440, in \S3 we discuss the data reduction method, in \S 4 and 
\S 5, we present the analysis of the $\rm \HI{}$ data cube and the results from the modeling of the three-dimensional structure of neutral hydrogen. 
In \S 6, we describe the optical photometry of FGC 1440. In \S 7 and \S 8, we present the results from mass models and constrain the vertical velocity dispersion using the observed scaleheight 
in conjugation with the best-fit mass model using the two-component Jeans equation. We finally present the associated discussion in \S 9 and conclude in \S 10.

\section{Target: FGC 1440}
FGC 1440 is an edge on disc galaxy included in the ultra flat galaxy catalogue \citep{karachentseva2016ultra}. Ultra flat galaxies are defined by very 
large axial ratio $(a/b)_{B}>10$, where $a/b$  is the ratio of the major axis to minor axis. The major and 
minor axes of FGC 1440 \citep{karachentsev1993flat, karachentsev2003revised} in $B$-band is $a_{B} \times b_{B}=2.24\farcm \times 0.11\farcm$ which gives an $(a/b)_{B}=20.36$.
FGC 1440 is an edge-on $(i=90^{\circ})$ late type Sd spiral galaxy \citep{de1991third}, located at a distance of  59.6 Mpc \citep{kourkchi2020cosmicflows}. 
In a study \cite{hoffman1989hi} report an $\rm \HI{}$ diameter $3.8\farcm$ and further indicate that the ratio of dynamical mass to blue luminosity($M_{dyn}/L_{B}$) 
is equal to 11.9 $M_{\odot}/L_{\odot}$, lower limit on the ratio of $\rm \HI{}$ mass to blue luminosity being $M_{\HI{} }/L_{B}>0.86 \, M_{\odot}/L_{\odot} $. 
The $\rm \HI{}$ properties of FGC 1440 have also been delineated in the ALFALFA $\rm \HI{}$ source catalogue, 
\cite{haynes2018arecibo} report $\HI{}$ flux $F=9.53 \pm 0.09$ Jy\,km\,s$^{-1}$ , $\rm \HI{}$ mass $log(M_{ \HI{} } )=10.08 \pm 0.18 M_{\odot}$,  and 
$W_{50}=298 \pm 2 \, \kms$. 
In their detailed study on the optical photometry of 47 late-type galaxies \cite{dalcanton2000structural, dalcanton2002structural} have observed that FGC 
1440 has an extremely small bulge in the center, but they conclude that it might not be a kinematic bulge but rather an edge-on orientation of pseudo-bulge. 
Further, from their study of vertical color gradients \cite{dalcanton2004formation} point out that the  FGC 1440 might also host concentrated dust lanes.
The studies investigating the kinematics of the thick disc in \cite{yoachim2008kinematics} find that FGC 1440 does not show signatures of a thick disc component, 
based on their measurement of the off-plane rotation curve, which they find is very similar to the mid-plane rotation curve. We have summarized the basic properties of 
FGC 1440 in Table 1.

\begin{table}
\begin{minipage}{110mm}
\hfill{}
\caption{Basic properties: FGC 1440}
%\centering
\begin{tabular}{|l|c|}
\hline
\hline
Parameter& Value \\
\hline    
RA(J2000)$^{\textcolor{red}{(a)}}$&  $12^{h}28\farcm52.29\farcs$  \\
Dec(J2000)$^{\textcolor{red}{(b)}}$& $+04{d}17\farcm35.4\farcs$  \\
P.A$^{\textcolor{red}{(c)}}$& $53^{\circ}$\\
a/b $^{\textcolor{red}{(d)}}$ & 20.4  \\
Hubble type $^{\textcolor{red}{(e)}}$ & Sd \\
$i$ $^{\textcolor{red}{(f)}}$ & $90^{\circ}$\\
Distance$^{\textcolor{red}{(g)}}$ &59.6 Mpc\\
log (M$_{HI}/M_{\odot} $)$^{\textcolor{red}{(h)}}$& 10.1\\
W$_{50}$ $^{\textcolor{red}{(i)}}$ & 298 \kms{} \\
D$_{HI}$ $^{\textcolor{red}{(j)}}$&3.8$\farcm$\\
M$_{HI}$/L$_{B}$ $^{\textcolor{red}{(k)}}$ & $>$ 0.86\\
M$_{Dyn}$/L$_{B}$ $^{\textcolor{red}{(l)}}$ & 11.8 $M_{\odot}/L_{B}$ \\
\hline
\end{tabular}
\hfill{}
\label{table: table 1}
\end{minipage}

\begin{tablenotes}
\item  $\textcolor{red}{(a, b, c, d)}$:  \cite{karachentsev2003revised} .
\item $\textcolor{red}{(e)}$:       \cite{de1991third}  .
\item $\textcolor{red}{(f)}$:       \cite{makarov2014hyperleda}. 
\item $\textcolor{red}{(g)}$:       \cite{kourkchi2020cosmicflows}. 
\item $\textcolor{red}{(h, i)}$:     \cite{haynes2018arecibo}.
\item $\textcolor{red}{(j, k, l)}$:   \cite{hoffman1989hi}.
\end{tablenotes}

\end{table}
\section{Observations and Data Reduction}
We observed FGC 1440 with the Giant Meterwave Radio Telescope (GMRT) on  August 19, 2019, for 7 hours (including the overheads) with 
26 working antennas. The target FGC 1440 was observed for 5.5 hours in 11 scans of 30 minutes each, interspersed by 11 
observations of phase calibrator 1150-003 of five minutes each. The flux calibrator 3C286 was observed at the beginning and the end of the observation for a
total of 30 minutes. The observation of the central frequency 1402.5 MHz was done in GSB mode with 512 channels with a spectral resolution of 8.14 kHz(1.71 \kms) 
and a total bandwidth of 4.14 MHz. Details of the observation are summarized in Table 2.
\begin{table}
\caption{Summary of FGC 1440 observation}
%\centering
\begin{tabular}{|l|c|}
\hline 
\hline
(a) Observing Setup\\
\hline
Parameter& Value \\
\hline    
\hline
Observing Date                     & 19August2019\\
Phase center,$\alpha$(J2000)       & $12^{h}28\farcm52.29\farcs$\\
Phase center,$\delta$(J2000)       & $+04^{d}17\farcm35.4\farcs$\\
Total on-source observation time   & 5 $\frac{1}{2}$ hours\\
Flux  calibrator                   & 3C286  \\
Phase calibrator                   & 1150-003\\
Channel Width                      & 8.14 kHz\\
Velocity separation                & 1.7 \kms\\
Central frequency                  & 1400.5 MHz\\
Total bandwidth                    & 4.14 MHz   \\
\hline
(b) Deconvolved Image Characteristics\\
\hline
Weighing                           & Briggs\\
Robustness parameter               & 0\\
Synthesized beam FWHM              & $15.9\farcs \times 13.5\farcs$\\
Synthesized beam position angle    & $23.1^{\circ}$\\
rms noise in channel               & 1.01 mJy/beam\\
\hline
\end{tabular}
\hfill{}
\label{table: table 2}
\end{table}

\subsection{Flagging and Calibration}
We perform the data reduction of our GMRT observation using Common Astronomy Software Application (CASA) \citep{mcmullin2007casa}. We begin by flagging the known bad antennae E04, E05, E06, and S02, which were offline during the observation. We then visually inspect the data set and flag the corrupted data. 
After flagging, we follow the usual procedures for cross-calibration. After cross-calibrating and splitting the 
$'target'$ from the measurement set, we average the visibilities in time to locate the channels containing the spectral line and
then flag those channels to create again a $'continuum-only'$ measurement set which we will use for self-calibration. 

\subsection{Imaging the spectral line}
We make our zeroth dirty image using CASA task $\sc{TCLEAN}$ and manually mask the emissions. We further do about four rounds of $'phase-only'$ self-calibration and three rounds of $\rm 'amplitude-phase'$ self-calibration,  
lowering the cleaning threshold in each round. We don't find any improvement in the rms value with further self-calibration and post continuum imaging. We do not detect any 
continuum emission from the center or around the galaxy FGC 1440. We further apply the final amplitude-phase self-calibration table to the $\rm 'target-only'$ measurement set 
containing both the spectral lines and continuum emissions before performing continuum subtraction using CASA task $\sc{UVCONTSUB}$ with zeroth order interpolation excluding
the spectral channels. We create a data cube with the task $\sc{tclean}$ and clean the emission within a mask made by the Source Finding Application 
(SoFiA, \cite{serra2015sofia}) down to a level of $0.5 \sigma$. We iterate this process until the SoFiA mask is stable. We have experimented with different weighting 
schemes in $\sc{TCLEAN}$, and we find that $\sc{briggs}$ weighing scheme with robustness parameter equal to zero and a uvtaper of $10k\Lambda$ gives us the best compromise between resolution and sensitivity. We finally  perform Hanning Smoothing on the cube and find that the final resolution of the data cube is $15.9\farcs \times 13.5\farcs$, 
and the rms noise is 1.01 mJy/beam compared to the expected theoretical noise is 1mJy/beam.

\section{Analysis}
\subsection{Global \HI{} profile}
\begin{figure}
\resizebox{60mm}{45mm}{\includegraphics{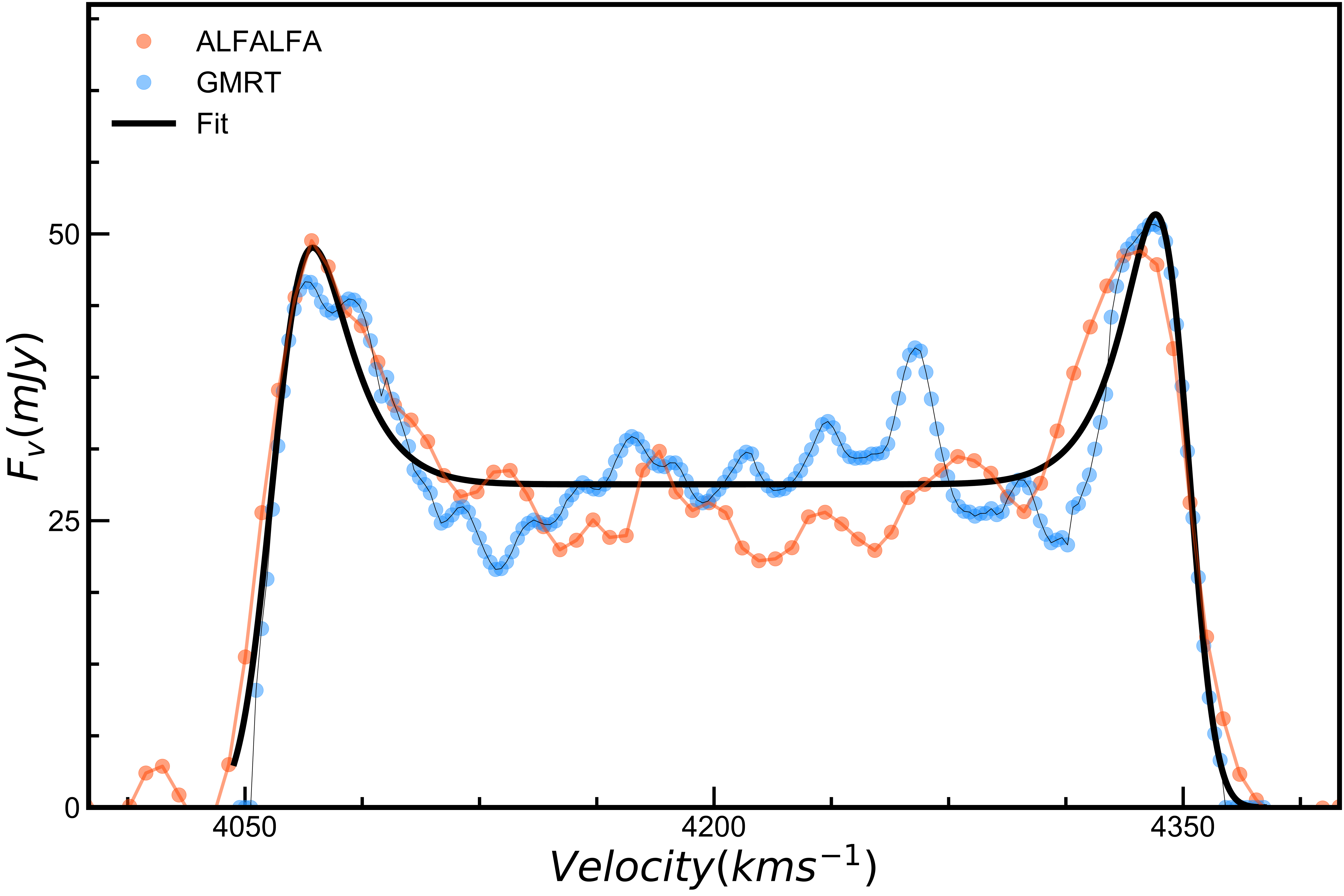}} 
\caption{Global HI profile of FGC 1440 derived from our GMRT observation. We have over plotted the fitted busy function on the observed spectrum and indicated
the derived profile parameters.}	
\end{figure}

In Figure 1, we present the global \HI{} profile of the galaxy FGC 1440. We note that our integrated HI flux $9.6$ $Jy\, \kms$ is comparable to that obtained in single dish observation by \cite{haynes2018arecibo}. We fit the observed profile using busy-function \citep{westmeier2014busy} to
derive the parameters corresponding to the profile. We find that the velocity widths 20$\%$ and the 50$\%$ of the peak maximum are 304.9 \kms and 293.3 \kms respectively. 
The results are summarized in Table 3.
\begin{table}
\begin{minipage}{110mm}
\caption{Best fit values obtained by fitting busy function.}
\centering
\begin{tabular}{|c|c|c|c|c|}
\hline
\hline
$V^{\textcolor{red}{(a)}}_{0}$   & $W^{\textcolor{red}{(b)}}_{50}$ & $W^{\textcolor{red}{(c)}}_{20}$&$F^{\textcolor{red}{(d)}}_{peak}$& $F^{\textcolor{red}{(e)}}_{int}$ \\
\kms{}&\kms{}& \kms{}& mJy & $Jy\, \kms{}$     \\
\hline
$4206 \pm 1.9$ &$293.3\pm 2.1$&$304.9\pm 2.94 $&$51.7 \pm 3.7$&$9.6 \pm0.2$\\
\hline
\end{tabular}
\hfill{}
\label{table: table 3}
\end{minipage}
\begin{tablenotes}
\item  $\textcolor{red}{(a)}$: Frequency centroid of the \HI{} line.
\item $\textcolor{red}{(b)}$:  Spectral line width at 50$\%$ of the peak flux density.
\item $\textcolor{red}{(c)}$:  Spectral line width at 20$\%$ of the peak flux density. 
\item $\textcolor{red}{(d)}$:  Peak of the \HI{} flux density.
\item $\textcolor{red}{(e)}$:  Integrated \HI{} flux.
\end{tablenotes}
\end{table}

\subsection{Channel and moment maps}
\begin{figure*}
\resizebox{150mm}{190mm}{\includegraphics{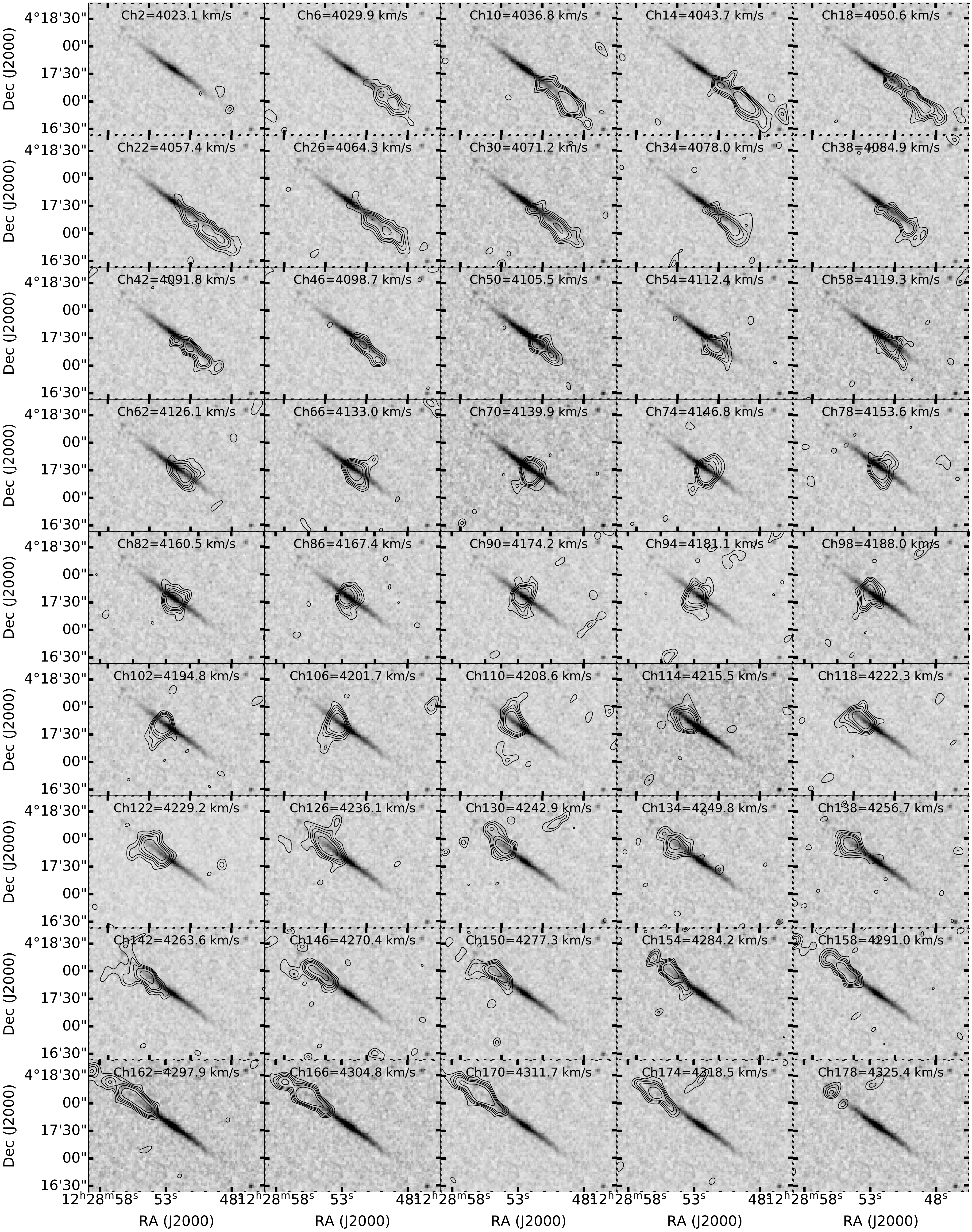}} 
\caption{Channel maps showing \HI{} emission from FGC 1440, each panel is separated by four channels. The \HI{} emissions are overlaid on the POSS-II 
optical image. The contour levels are at [3, 4, 5 ,6]$\times$ 1.01 mJy beam$^{-1}$}	
\end{figure*}

In Figure 2, we show the channel maps containing the \HI{} emission overlaid on POSS-II (The Palomar Observatory Sky Survey) optical images. 
In the channels with the highest deviation from the systemic velocity, the emission starts to emanate from the edge of the stellar disc and extends further beyond it.
In some channels close to the systemic velocity (70, 102, 106, 110), we see the  \HI{} emission extends out of the plane at the center, as compared to the channels
further away from the center, for example, the emissions from channels 70, 102 are remarkably extended compared to channels 170, 174, or channels 14 and 18. 
We will discuss in detail the possible origins of thickening observed in the central channels in \S 5.4 

\begin{figure}
\resizebox{90mm}{85mm}{\includegraphics{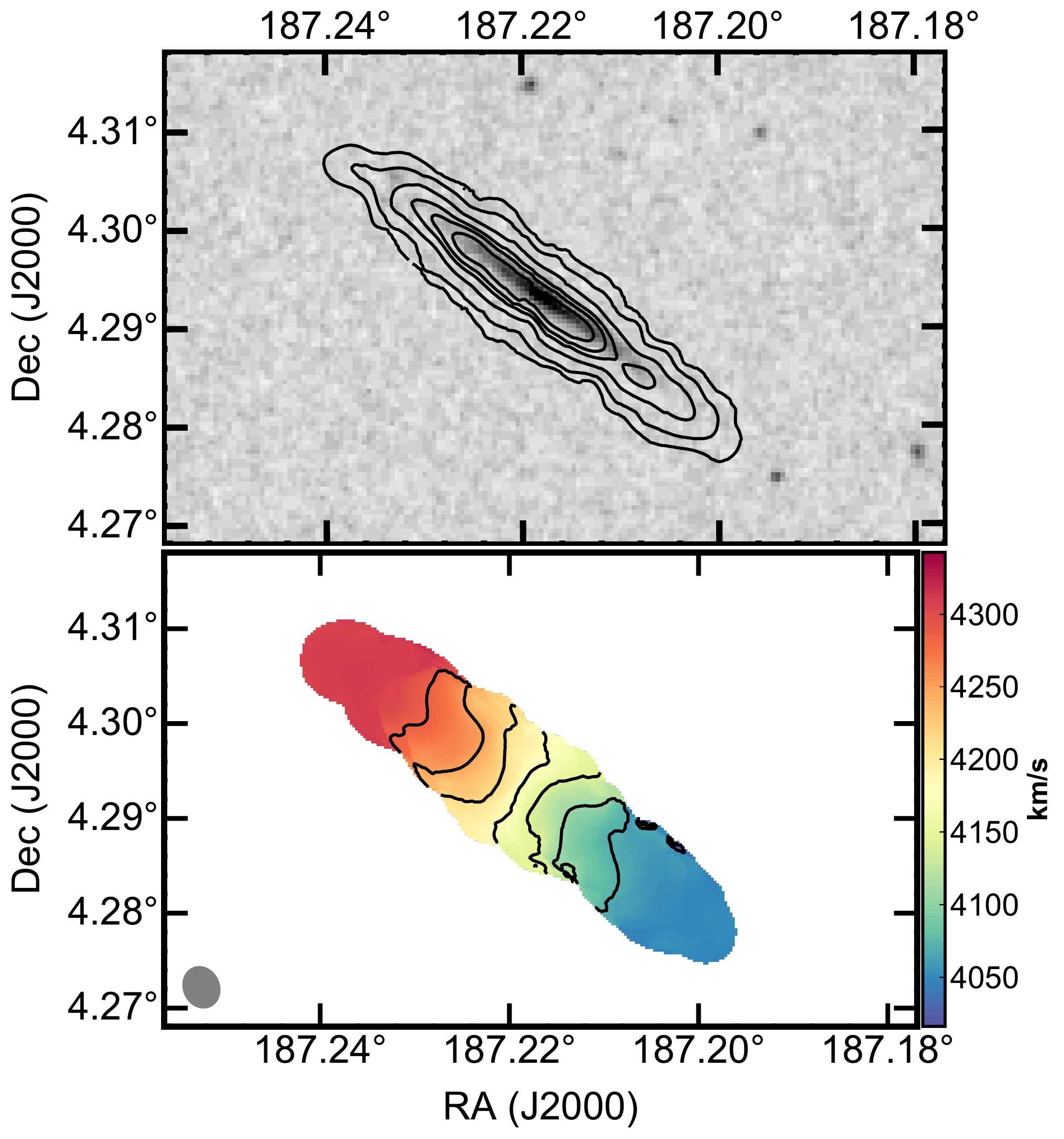}} 
\caption{In the top panel we have plotted the Moment0 map the contours are at [ 2.5, 5.0, 10, 15, 18, 22]$\times$ 40 mJy beam$^{-1}$ kms$^{-1}$. 
In the bottom panel, we have shown the Moment 1 map, and the contours start at 4000 kms$^{-1}$ increasing at 35 km s$^{-1}$. }
\end{figure}

In Figure 3, we have shown the Moment 0 and the Moment 1 map. We have overlaid the Moment 0 map on the POSS-II optical image. From Moment 0 map,
we see \emph{a slight warp on the North-East side of the galaxy}.

\section{3D - Tilted Ring Modeling}
We use the publicly available tilted ring modeling software TiRiFiC \citep{jozsa2012tirific} for deriving the kinematic and structural properties of FGC 1440. 
We model the galaxy as a rotating gas disc of radius R with the gas vertical distribution following $\rm sech^{2}$ profile. 
Each ring is set to 1.1 (5.1 \,kpc) times the beamwidth.

\subsection{Modeling strategy} For estimating the initial fit we use Fully Automated TiRiFiC \citep[FAT,][]{kamphuis2015fat} which is GDL wrapper 
around TiRiFiC. We then further use these initial estimates of the fit parameters and compare the model and data by \textbf{(1)} visually inspecting the emissions in individual channels, \textbf{(2)} comparing the Moment 0 and Moment 1 maps, and \textbf{(3)} the PV diagrams at various offsets to check how well do the model contours match the data. 
We then further manually vary individual parameters using TiRiFiC while comparing the model and data visually in each iteration. We will now discuss how each 
kinematics and the structural parameters describing the tilted ring model describing FGC 1440 were chosen.

\subsection{Automated Fit using FAT} We have used the beta version of FAT which allows for radial variation of the intrinsic velocity dispersion. 
FAT takes an \HI{} data cube as an input and automatically estimates the following parameters (free-parameters); 
1)surface brightness profile, 2)  position angle 3) inclination 4) rotational velocity 4) scaleheight 5) intrinsic 
dispersion and 5) central coordinates:- right ascension, declination and the systemic velocity. FAT fits each parameter ring by ring and finally smooths the 
parameters with a polynomial of order 0, 1, 2, 3, 4 or 5.  FAT models the \HI{} disc as two halves and fits 9 semi-rings across each half. 

The procedure employed by FAT for finding the initial estimates of the 
free parameters and the flow-chart for fitting is discussed in detail in \cite{kamphuis2015fat}.
FAT fits the surface brightness of the approaching side and the receding side independently and adopts scaleheight equal to $6.1\farcs$.  
We have summarized the model parameters estimated by FAT in Table 4 and present the plots Figure 4. The rotation velocity $(V_{Total})$ derived using 3D tilted ring modeling
is shown in Figure. 5 overlaid on a major axis PV diagram.
After getting an estimate of the kinematic 
and structural parameters using FAT, we investigate how sensitive the FAT-model is to the variation of each parameter by manually running TiRiFiC, 
each time changing the fit parameters obtained from FAT and then comparing the model with the data. We find that the model data cube is  degenerate
for a large range of dispersion values, scale height, and inclination. 

\begin{figure}
%\vspace{-5mm}
\hspace{-10mm}
\resizebox{100mm}{95mm}{\includegraphics{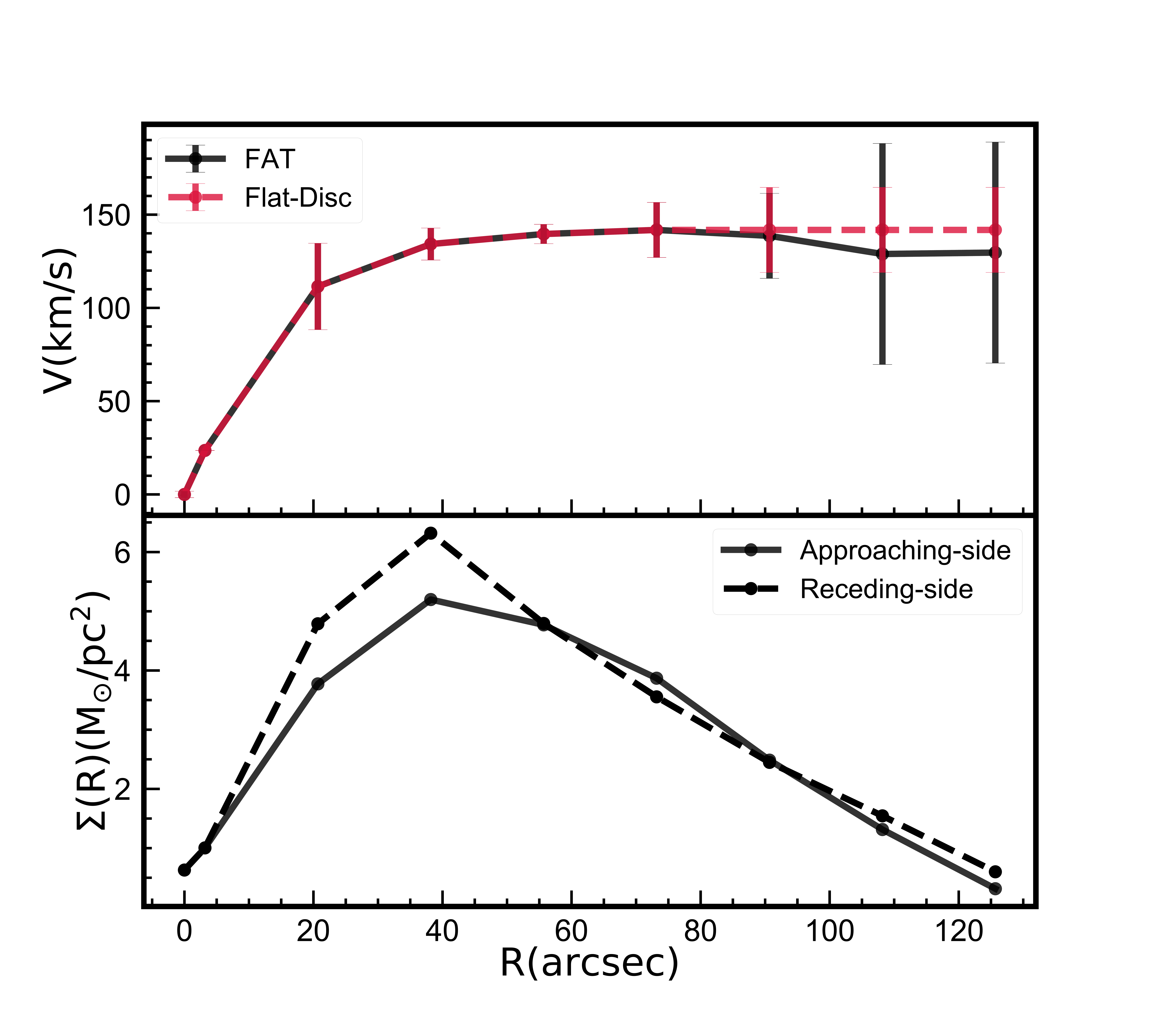}} 
%\vspace*{-15mm}
\caption{In the above figure we have plotted the rotation velocity and the surface brightness profile obtained from the tilted ring modeling.
The surface brightness is fitted independently for the approaching and the receding side.}	
\end{figure}

\begin{figure}
\hspace{-10mm}
\resizebox{100mm}{49mm}{\includegraphics{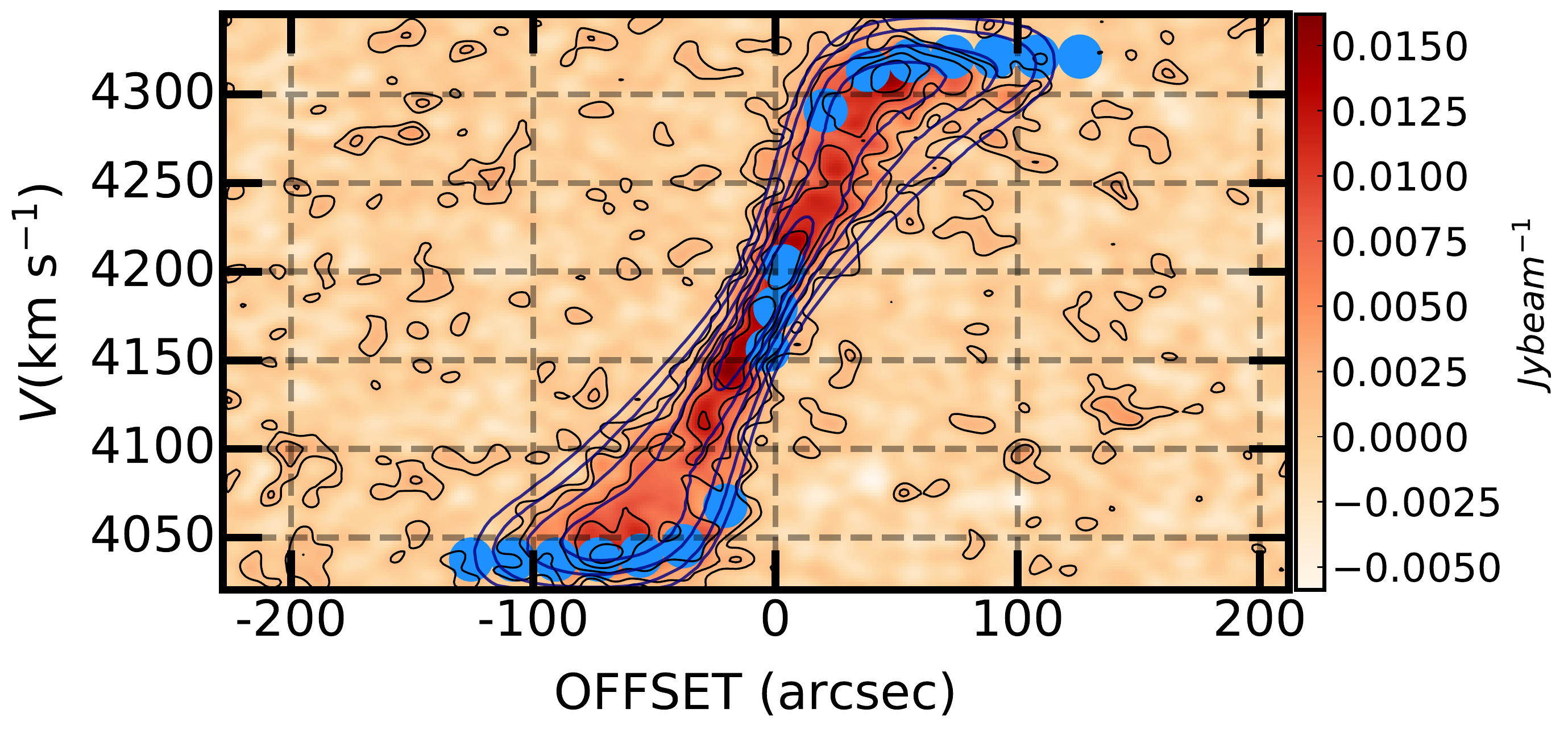}} 
\caption{The major axis PV map with flat disc model overlaid at contours [1.5, 3, 6, 9, 12]$\times$ 1.01 Jy/beam. The blue points describe the \HI{} rotation curve.}
\end{figure}

\begin{table}
\begin{minipage}{110mm}
\hfill{}
\caption{Parameters describing 'FAT' and 'Flat-Disc' models.}
%\centering
\begin{tabular}{|l|c|c|}
\hline
\hline
Parameter&    FAT-Model& Flat-Disc-Model  \\
\hline    
$\rm X_{o}$\footnote{Right ascension}            &    $+187.2$                &   $+187.2$  \\
$\rm Y_{o}$\footnote{Declination}            &    $+4.29$                  &   $+4.29$    \\
$\rm i$ \footnote{Inclination}               &    $85^{\circ}$                  &   $88.5^{\circ}$        \\
$\rm V_{sys}$\footnote{Systemic velocity}          &    4179.45                &   4179.45     \\
PA \footnote{Position angle}                    &    $53.5^{\circ}$         &   $53.5^{\circ}$     \\
Surface Brightness \footnote{Surface brightness as a function of radius}    &    fig 4                   &   fig 4            \\
Dispersion \footnote{Velocity dispersion}             &    6.6\kms               &   15 \kms\\  
Rotation velocity \footnote{Rotation velocity as a function of radius}     &    fig 4                   &   fig 4             \\
$\rm h_{z}$ \footnote{Scaleheight of the \HI{} disc}           &    6.10 \farcs (1.8 kpc)             &   0.45 \farcs (0.13 kpc)  \\
\hline
\end{tabular}
\hfill{}
\label{table: table 4}
\end{minipage}
\end{table}

\subsection{Manual TRM Models}
In an attempt to break the degeneracy between the parameters, by constructing a $Flat-Disc$ model, We perform several rounds of manual fitting with TiRiFiC followed 
by visual inspection of channel maps, PV diagrams at various offsets, and the moment maps to estimate the parameters describing the $Flat-Disc$ model. 
The method of iterative visual inspection to routinely carried out to fine-tune the parameter and arrive at the final model data cube see 
for example \cite{allaert2015herschel} and also \cite{zschaechner2012halogas}, \cite{gentile2013halogas}, \cite{kamphuis2013halogas}. 

Using the model data cube output by FAT as the base model, we iteratively construct model data cubes, each time changing the following parameters; 
inclination $(i)$, dispersion $(\sigma)$, and the scaleheight$(h_{z})$ in the model output by FAT. In the $Flat-Disc$ model, we fix the center's values (RA, Dec), 
systemic velocity, surface brightness, and position angle to the values derived by FAT. We make different model data cubes by  varying the 
parameters $(i,\, \sigma,\, h_{z})$ either one by one, in pairs or all together at the same time.
We compare each model with the data through a visual-inspection, wherein we compare the model and data channel-wise, comparing 3D 
models of data and the model data cubes using volume rendering software astroslicer \citep{punzo20173d}, and comparing the datacube 
by taking slices at various offsets along the major axis (as it preserves the 3D structure of the cube) to arrive at the secondary base model 
called the $'Flat-Disc'$ model. In Figure 6, we show the Moment 0 and the Moment 1 map derived for the 'flat-disc model' superposed on the data. 

In order to show the effect of varying the inclination $(i)$, dispersion $(\sigma)$ and the scaleheight $(h_{z})$, 
we plot the minor axis PV diagrams (Figure 7 to 11) at different offsets,
varying each of the parameter $(i, \, \sigma, \,h_{z},)$ one by one, while keeping the values of the other two-parameters to
that of the  $Flat-Disc$ model (Table. 4). For example, in the second row of the Figure 6, we show the effect of variation of the inclination and how it compares with the
data. We keep the values of the dispersion and scaleheight equal to that of the $Flat-Disc$ model and vary the values of the inclination. Similarly, in the third row, we keep 
the values of inclination and scaleheight fixed to that of $Flat-Disc$ and vary only the dispersion.

Also, we consider the models in which we vary the values of the inclination, dispersion, and
scaleheight as a function of radius, keeping the value of other parameters to be same as that of the $'Flat-Disc'$ model;
\begin{itemize}
\item \textbf{Radially inclination $i(R)$:} The inner rings are kept at an inclination equal to 90$^{\circ}$ and the outer rings at 85$^{\circ}$.\\
\item \textbf{Radially varying dispersion $\sigma(R)$:} Dispersion varies from 20 \kms{} for the inner rings to 5 \kms{ at the outer rings in steps of 5 \kms{}}.\\
\item \textbf{Radially varying scaleheight$h_{z}(R)$:} The inner rings are kept at an $h_{z}=0.45\farcs{}$, the central rings are kept at $h_{z}=1.97\farcs{}$, and the outer rings at $h_{z}=5.3\farcs$
\end{itemize}

\textbf{Inclination $(i)$.} For studying the effect of inclination, we fix the values of all the parameters to that of the 'flat disc model' and vary just the inclination to different values. 
From the minor axis PV diagrams (see Figure 7 to Figure 11), the first immediate observation is that we can rule out the models with an $i<85^{\circ}$, 
for example, the PV plot at an offset equal to 0, Figure 7, the inner model contours at 9$\times$rms are not extended sufficiently to describe the emissions and, 
further models with lower inclination only increase this discrepancy. Comparing the PV diagrams at different offsets, we find that the inclination is restricted in the range $85^{\circ}<i<90^{\circ}$. We note that the inclination value estimated by FAT $i=85^{\circ}$ is the lower limit for the inclination of FGC 1440. 
Models with an inclination lower than $85^{\circ}$ do not describe the data accurately.

\textbf{Dispersion $(\sigma)$.} The value of dispersion is set to 15 \kms, whereas FAT fits dispersion profile; 6.63 \kms. By comparing the PV diagrams 
(figure 7 to 11) we find that the data is not very sensitive to the change in dispersion as the models with dispersion 5 \kms to 15 \kms show little variation at the level 
of data. However, we find that the models with dispersion greater than 15 \kms clearly start to deviate from the data. We find that the value of the 
dispersion describing the data is restricted in the range $5\kms<\sigma_{z}<15 \kms$.

\textbf{Scaleheight $(h_{z})$.} In Fig. 6 - 9, we observe that the models with higher scaleheight are spatially more extended as compared to models with lower 
scaleheight, which is more extended along the velocity axis but not spatially.
Further, we observe that, in the case of models with higher scaleheight, 
the model contours do not follow the data contours in the inner regions, i.e., the contours at $9\times rms$. The models with lower scaleheight do
follow the data contours in the inner regions of the PV diagrams, but it is not easy to distinguish between the models with $h_{z}=0.45\farcs{}$ and 
$h_{z}=1.97\farcs{}$. We further construct   
a model with radially varying scaleheight, based on the observation that the inner rings at lower scaleheight give a better description of the data contours in the innermost region of the PV diagram, and the outer rings kept at the higher scaleheight will match the spatially extended data contours in the outer region of the PV diagram.
We find the model with radially varying scaleheight is barely distinguishable from models with $h_{z}=0.45\farcs{}$ or $h_{z}=1.97\farcs{}$. 
To further investigate the vertical structure, in figure 12, we plot the normalized vertical density profile extracted at various slices from
the moment 0 maps and overlaid the major axis FWHM of the synthesized beam. We observe that the synthesized beam is 
comparable to the vertical density profile extracted from the data, thereby indicating that the disc's thickness is barely resolved in these observations.
From comparing PV diagrams, we find that the upper limit on scaleheight is 5.3 \farcs{}.

By comparing the model and data in the minor axis PV diagrams  Figure 7 - 11, following above discussion, we find the lower
and upper limits for the  values for the inclination $(85^{\circ}\leq i \leq 88.5^{\circ})$, dispersion $(5\kms \leq\sigma \leq 15 \kms)$ 
and the scaleheight$( h_{z} \leq 5.3 \farcs{} )$.

\subsection{Thickness of the \HI{} disc}
Is it a Flare, thick disc or a line of sight warp ?\\
In Figure 2, from the channel maps, we observe that the emissions from the channels close to the systemic velocity extend above the plane as 
compared to the emissions from the end channels, possibly indicating the presence of a thick \HI{} disc or a line of sight warp. We can possibly
rule out a flaring disc. The flare would be circular and hence every where along the line of sight. The flare would be better noticeable in the outer channel
because the sight line through the outer parts of the disc would be longer and hence the flare is harder to identify in the center. 
In order to investigate the origin of the observed thickness of the $\rm \HI$ disc in the FGC 1440, we make the Moment 0 map considering only the 
starting channels(4023 \kms - 4100 \kms), the central channels close to systemic velocity( 4100 - 4242 \kms) and the end channels (4242 -4325 \kms). 
We then compare the data contours for the said velocity ranges with the model with radially varying scaleheight, inclination, and the 
flat disc model. We find that the data contours (see Figure 13) are as thin as the model contours in high-velocity channels, indicating that 
possibly the \HI{} disc in FGC 1440 doesn't show flaring behavior. Further absence of flaring is supported by the observation that instead of 
the end channels at higher velocities, channels close to the central velocity contribute to the thickness of the \HI{} disc. Thus, a thick \HI{}
disc at the center with thickness tapering off radially leaves us only with two options; \emph{that either we are observing a line of 
sight warp or a thick HI disc at the center}. 

In order to further investigate this effect we consider the following models 1) A model with 
radially varying inclinations and 2) A two-component model with a lagging thick disc $(\frac{dv}{dz}=-10 \kms\, kpc^{-1})$ and $(h_{z}=5.3\farcs)$  
and a thin disc $(h_{z}=0.45\farcs)$. The value of scaleheight of the thick disc in the two-component model is equal to the upper limit on the 
scaleheight obtained for the one compoent model. We then iteratively compare models with data through visual-inspection for different values of $\frac{dv}{dz}$, and find that 
$\frac{dv}{dz}=-10 \kms\, kpc^{-1}$ better reproduces the data contours. From Figure 13 and 14, we find that overall models with radially varying inclination and the two-component model are very similar; it is barely possible to distinguish if 
the observed central thickness is the result of radially varying inclination or if the galaxy host central thick \HI{} disc.  Further, in Figure.12, we
compare the synthesized major axis beam with the density profile, indicating that we possibly can not resolve the thick disc in our observations, 
although we can not rule out the presence of a central thick \HI{} disc. We adopt a model with a line of sight warp as it is 
simpler model than the two-component model.

\begin{figure*}
\resizebox{160mm}{60mm}{\includegraphics{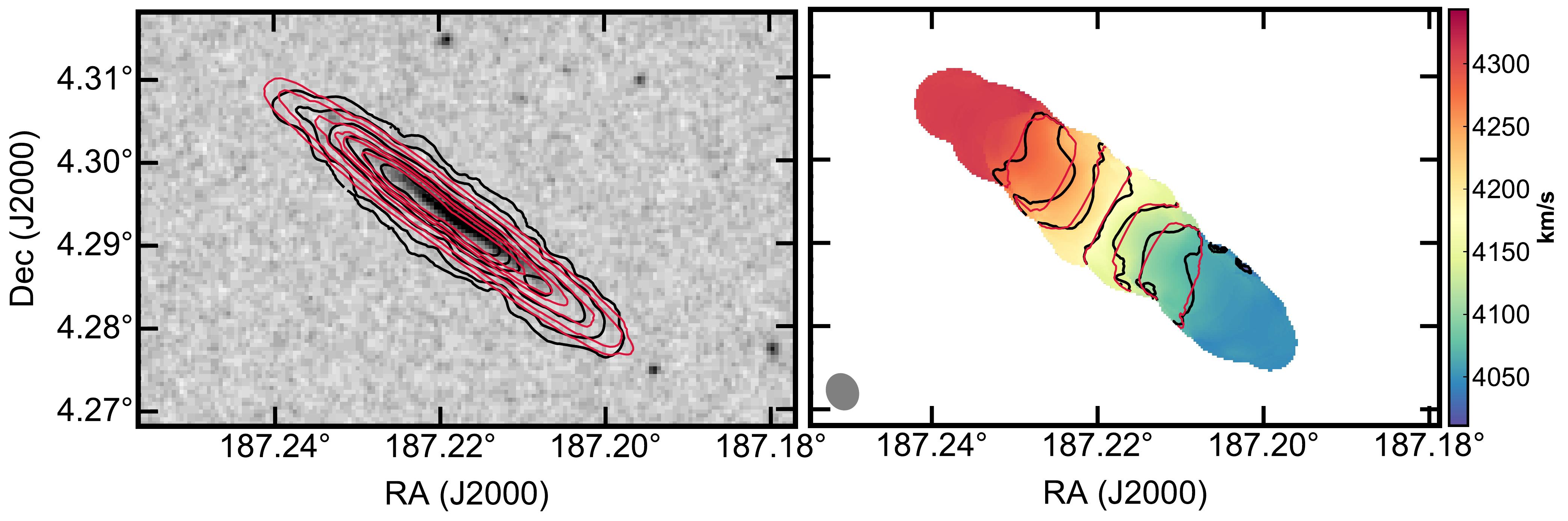}} 
\caption{In the top panel we have plotted the moment 0 (top) and the moment 1 (bottom) map the contours are at [2.5, 5.0, 10, 15, 18, 22]$\times$ 40 mJy beam$^{-1}$ kms$^{-1}$. 
and the contours for moment 1 map start at 4000 kms$^{-1}$ increasing at 35  km s$^{-1}$ respectively. The contours corresponding to the data are rendered in black 
and the contours $Flat-Disc-Model$ are in shown in red color.}
\end{figure*}

\begin{figure*}
\resizebox{160mm}{100mm}{\includegraphics{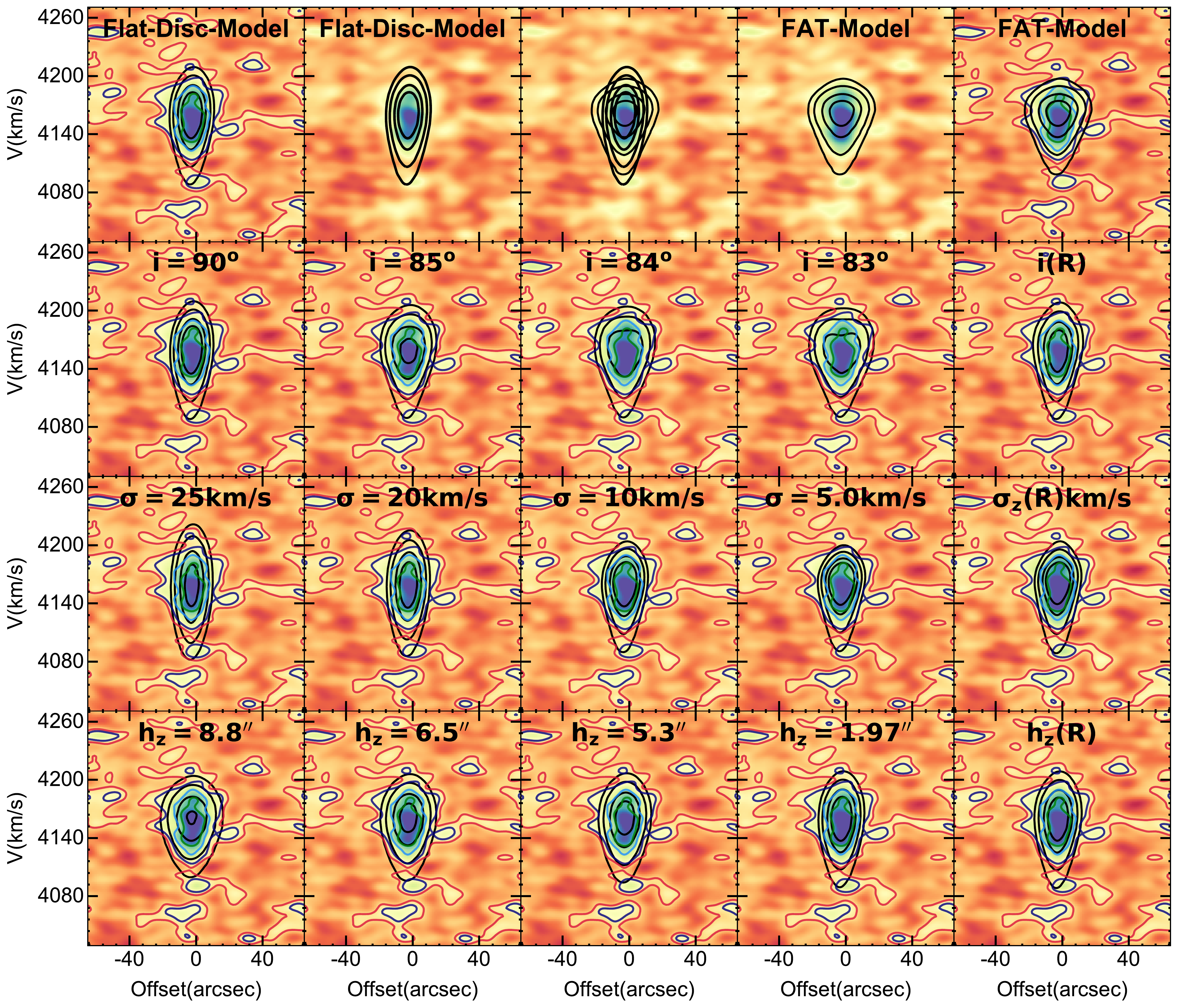}} 
\caption{Position velocity maps parallel to the the minor-axis comparing the tilted rings model to data at offset equal to 0 by varying the model
parameters. The contours at [1.5, 3, 6, 9]$\times$ 1.01 Jy/beam.}	
\end{figure*}

\begin{figure*}
\resizebox{160mm}{100mm}{\includegraphics{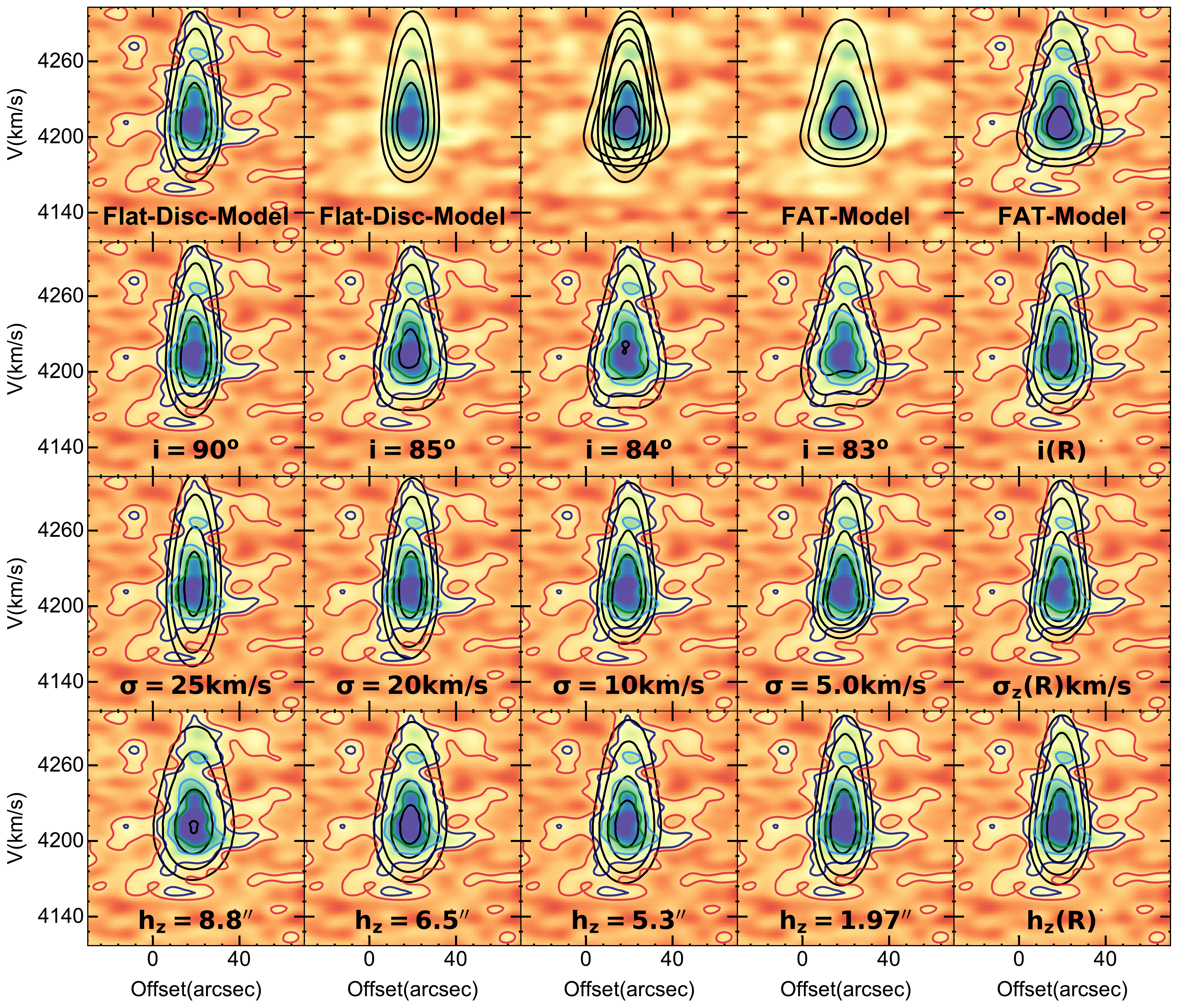}} 
\caption{Same as figure 6 but at offset=20}	
\end{figure*}

\begin{figure*}
\resizebox{160mm}{100mm}{\includegraphics{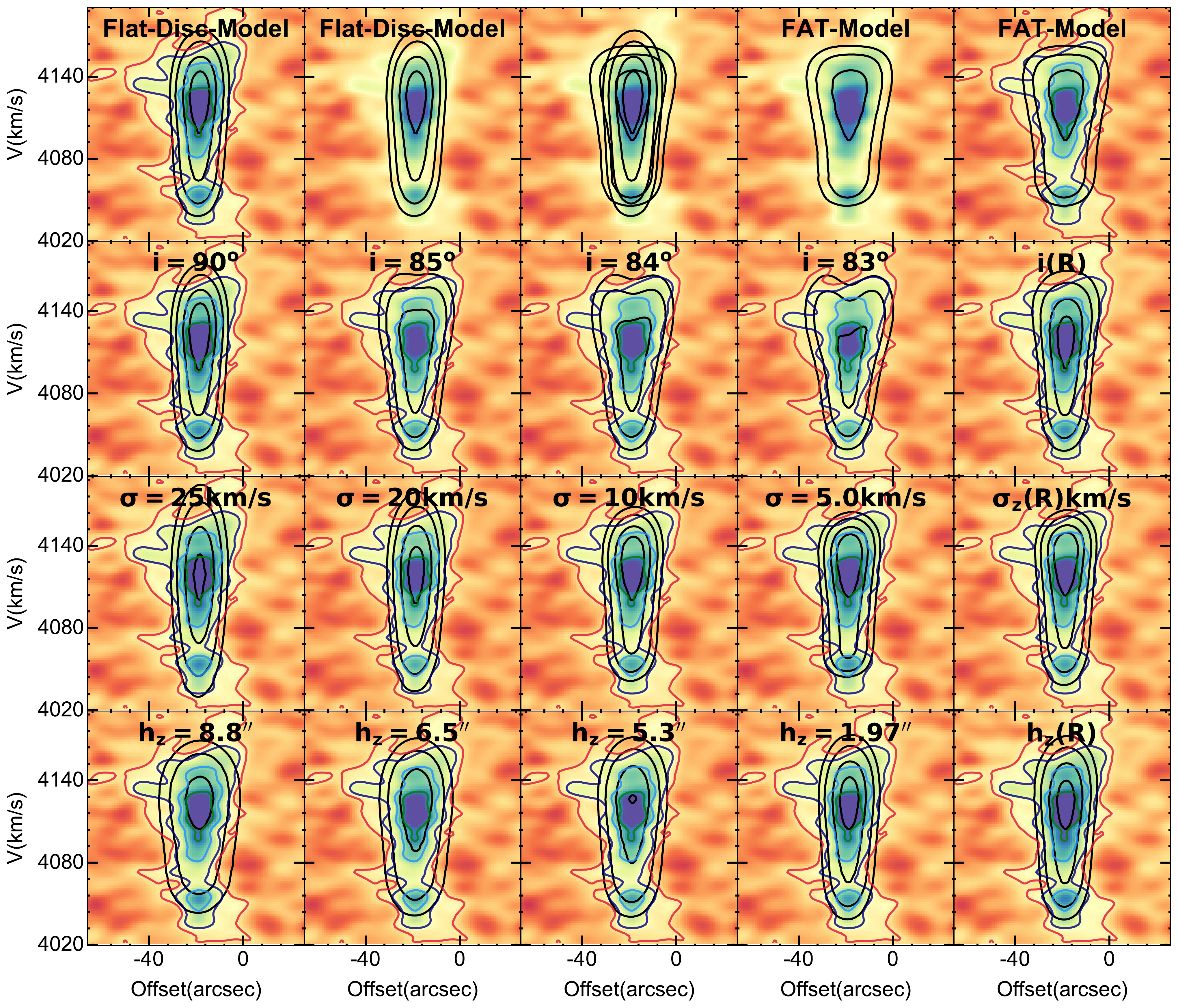}} 
\caption{Same as figure 6 but at offset equal to -20,.}	
\end{figure*}

\begin{figure*}
\resizebox{160mm}{100mm}{\includegraphics{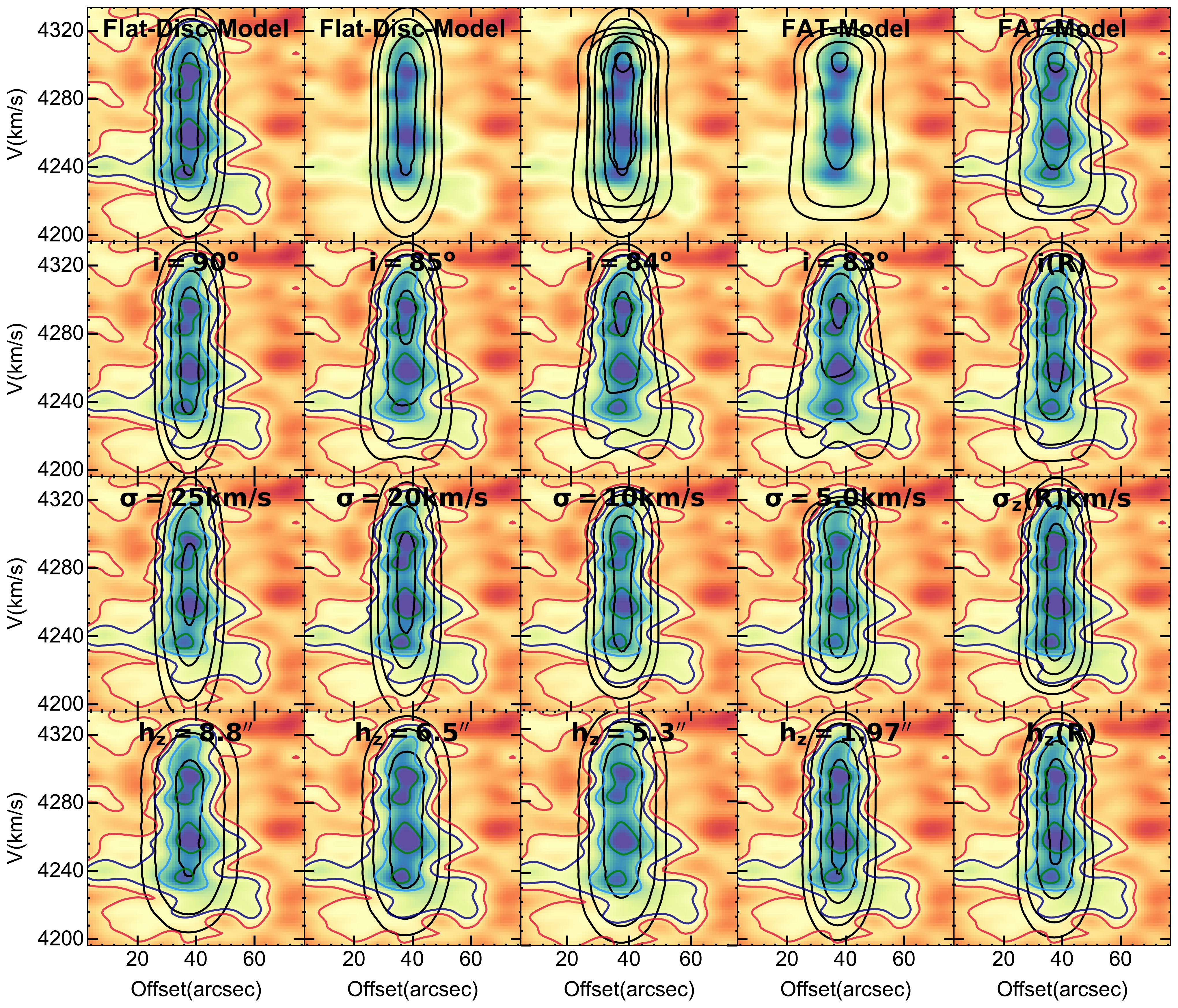}} 
\caption{Same as figure 6 but at an offset equal to 40.0 }	
\end{figure*}

\begin{figure*}
\resizebox{160mm}{100mm}{\includegraphics{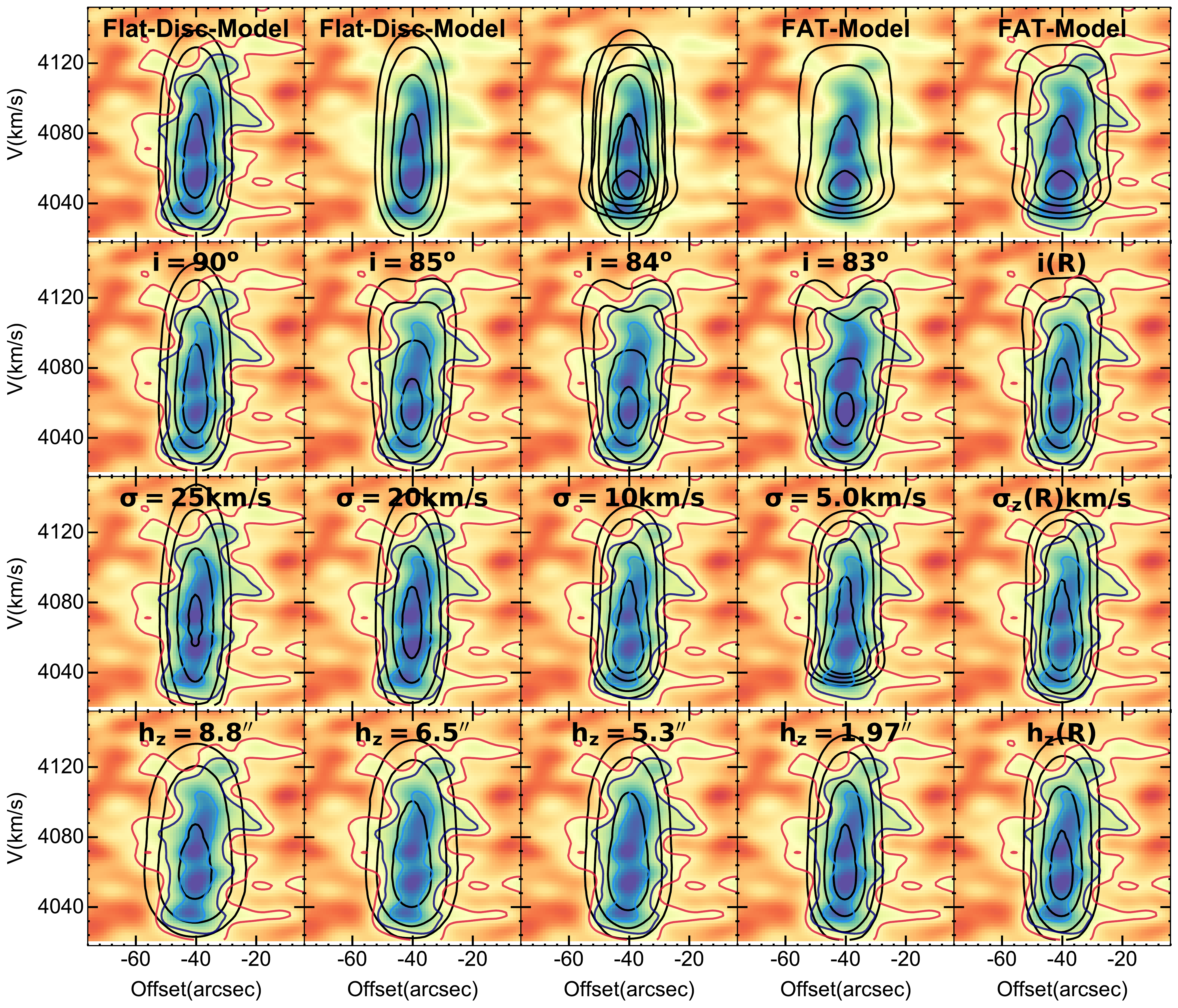}} 
\caption{Same as figure 6 but at offset equal to -40.}	
\end{figure*}

\begin{figure*}
\resizebox{180mm}{50mm}{\includegraphics{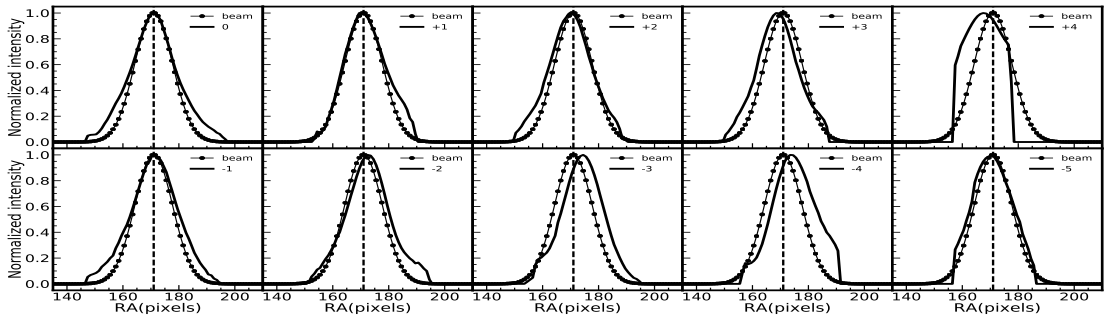}} 
\caption{We have compared the synthesized major axis beam size with the vertical density profile at slices extracted from the moment 0 maps.}	
\end{figure*}

\begin{figure*}
\resizebox{180mm}{80mm}{\includegraphics{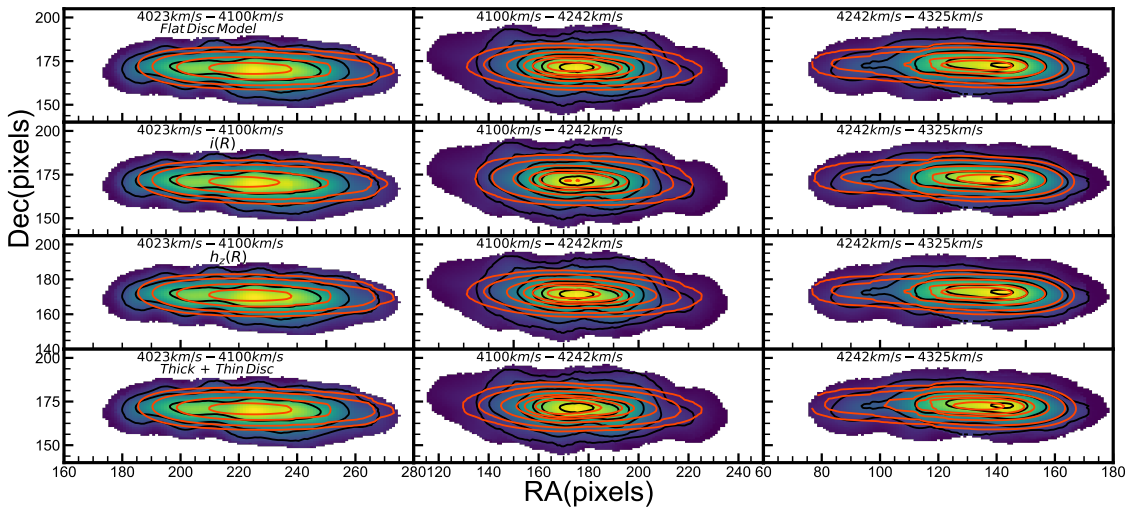}} 
\caption{We compare the moment zero maps derived from the central and end velocity channels. In first row we compare the moment maps for the $'Flat-Disc-Model'$
derived at the velocity range 4023 \kms to 4100 \kms the first column and in the middle panel moment 0 map for the central velocity range 4100 \kms - 4242 \kms, and in the 
third panel we show the moment 0 maps for the velocity range 4242 \kms to 4325 \kms. Similarly, in the second, third row, and  
fourth row, we have plotted the contours for the models with radially varying inclination, varying scaleheight, and the two-component model. }	
\end{figure*}

\begin{figure*}
\resizebox{180mm}{30mm}{\includegraphics{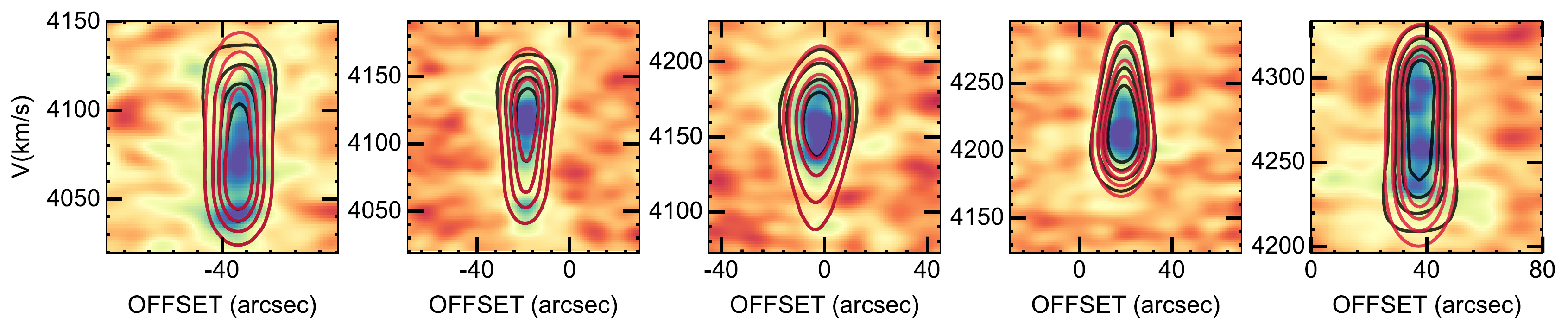}} 
\caption{We show the minor axis PV diagrams comparing the model with radially varying inclination and the two-component model. 
Black contours depict a model with radially varying inclination, and the red contours show the two-component model. }	
\end{figure*}

\section{Optical photometry}
In this section, we present the photometric analysis of FGC 1440 in SDSS g, r band, and UKIDSS K band. As a first step, we mask all the surrounding 
objects and the galaxy itself and estimate the positions and the positional angles (PA) using SExtractor \citep{bertin1996sextractor}. 
We subtract the background by fitting it with a two-dimensional second-order polynomial and then further rotate the entire frame by the PA.
Next, we eliminate all the objects lying close to the galaxy by replacing them with the regions symmetric to the galaxy's mid-plane. 
We then further integrate the light in a rectangular box to estimate the total magnitude. The size of the box was chosen in such a way 
that the growth curve becomes flat near the borders of the box. The estimated value of the total magnitude in the g, r, and K bands 
are 15.43, 14.72, and 11.83. We derive the structural parameters of the galaxy using GalFit 
\citep{peng2011galfit}. The intensity profile is $I \sim R/R_{d} K_{1}(R/R_{d})sech^{2}(z/h_{z})$, where $R_{d}$ is the disc scale radius, 
$h_{z}$ is the disc scaleheight, and $K_{1}$ is the modified Bessel function of the second kind. 
The parameters derived using optical photometry are summarized in Table 5. The data, the galfit model, and their normalized difference are shown in 
shown in appendix A.

\section{Mass Modeling}

\begin{table*}
\hspace{-6cm}
\begin{minipage}{110mm}
\hfill{}
\caption{Input parameters for deriving the stellar rotation curve.}
\centering
%\small\addtolength{\tabcolsep}{-1pt}
\begin{tabular}{|l|c|c|c|c|}
\hline
\hline
Parameter                                 &Value        &Value     &value            &Description \\

                                          &g-band&     r-band      &K-band           &             \\
\hline
Total magnitude                           &15.4&      14.7        &11.8           &             \\
$\mu^{edge-on}_{o}$                       &21.6&      20.8        &16.5           &Edge-on surface brightness in units of $mag/arcsec^{2}$\\
$\mu^{face-on(\textcolor{red}{*})}_{o}$   &23.3&      22.5        &18.6           &Face-on surface brightness in units of $mag/arcsec^{2}$\\
$\Sigma_{o}$                              & 22.7&      31.0        &328.0           &Surface density in units of $L_{\odot}/pc^2$\\
$R_{d}$                                   & 4.4&      4.2        &2.6            &Disc scalelength in units of kpc\\
$h_{z}$                                   & 0.9&      0.9          &0.4           &Disc scaleheight in units of kpc\\
\hline
Parameters for deriving  mass to light ratio $(\gamma^{*})$&  &    & \\
\hline
$g-r$         &0.7      &        &                     \\
$a_{\lambda}$ & -0.5&  -0.31      &-0.2  &\cite{bell2003optical}\\
$b_{\lambda}$ & 1.5&     1.1       &0.2   &\cite{bell2003optical}\\
$\gamma^{*}$  & 3.8&     3.0        &0.8   &M/L ratio derived using scaled Salpeter IMF\\
$\gamma^{*}$  & 2.7&     2.1         &0.6   &M/L ratio derived using Kroupa IMF\\
\hline
\end{tabular}
\hfill{}
\label{table: table 5}
\end{minipage}
\begin{tablenotes}
\item  \textcolor{red}{(*)}: The edge-on surface brightness has been converted to face-on surface brightness using 
$ \mu^{face-on}= \mu^{edge-on} + 2.5log( \frac{ R_{d} } {h_{z}})$ \citep{kregel2005structure}.
\end{tablenotes}
\end{table*} 
In this section we will present our analysis and results from mass models of FGC 1440. By decomposing the total rotation curve 
of the galaxy into baryonic (stars+$\rm \HI$) and dark matter components we will determine the contribution of each mass - component to the total rotation curve
$(V_{Total})$ (see Figure 4 ). The total rotation curve of the galaxy is obtained by adding in quadrature the 
circular velocity profiles produced by each component separately.

\begin{equation}
V^{2}_{Total}=\gamma V^{2}_{*} + V^{2}_{gas} + V^{2}_{DM}
\end{equation}
where $\gamma$ is the mass to light ratio (M/L), $V_{*}$, $V_{gas}$ and $V_{DM}$ are the circular velocity profiles due to 
the stars, gas and dark matter components respectively.

\subsection{Neutral gas distribution and rotation curve}
We model the gas disc as thin concentric rings and use GIPSY \citep{van1992groningen} task ROTMOD to derive $V_{gas}$. 
We use the surface densities as a function of radius obtained from the tilted ring model (panel 2 Figure 4) as the input parameter in ROTMOD.
The gas surface density is scaled by a factor of 1.4 to account for helium and other metals. 

\subsection{Stellar distribution and rotation curve}
For ascertaining the contribution of the stars to the observed rotation curve, we model the stellar distribution in the optical $g$ band and the near-infrared (NIR) UKIDSS K band.
We derive mass to light ratio $(\gamma^{*})$ ratio following the empirical relations in \cite{bell2003optical} and \cite{bell2001stellar} based on stellar 
population synthesis models. The scaling between the color magnitude and the mass to light ratio $(\gamma^{*})$

\begin{equation}
\gamma^{*}=10^{(a_{\lambda} +b_{\lambda} \times Color )}
\end{equation}

In the above equation $\gamma^{*}$ is the mass to light ratio, $a_{\lambda}$ and $b_{\lambda}$ are the intercept and slope of the 
$log_{10}(\gamma^{*})$ versus color calibration obtained by \cite{bell2001stellar} using stellar population synthesis models.
We compare the mass models derived using $'diet'$ Salpeter IMF with the mass models derived using Kroupa IMF \citep{kroupa2001variation}.
The $\gamma^{*}$ ratios assuming Kroupa IMF are derived by subtracting 0.15 dex from the constant term $a_{\lambda}$(table 5). 
We use the input parameters described in table 5 to derive the stellar rotation curve using the GIPSY task ROTMOD.

\subsection{H$\alpha$ Rotation curve}
\begin{figure}
\hspace{-1cm}
\resizebox{85mm}{60mm}{\includegraphics{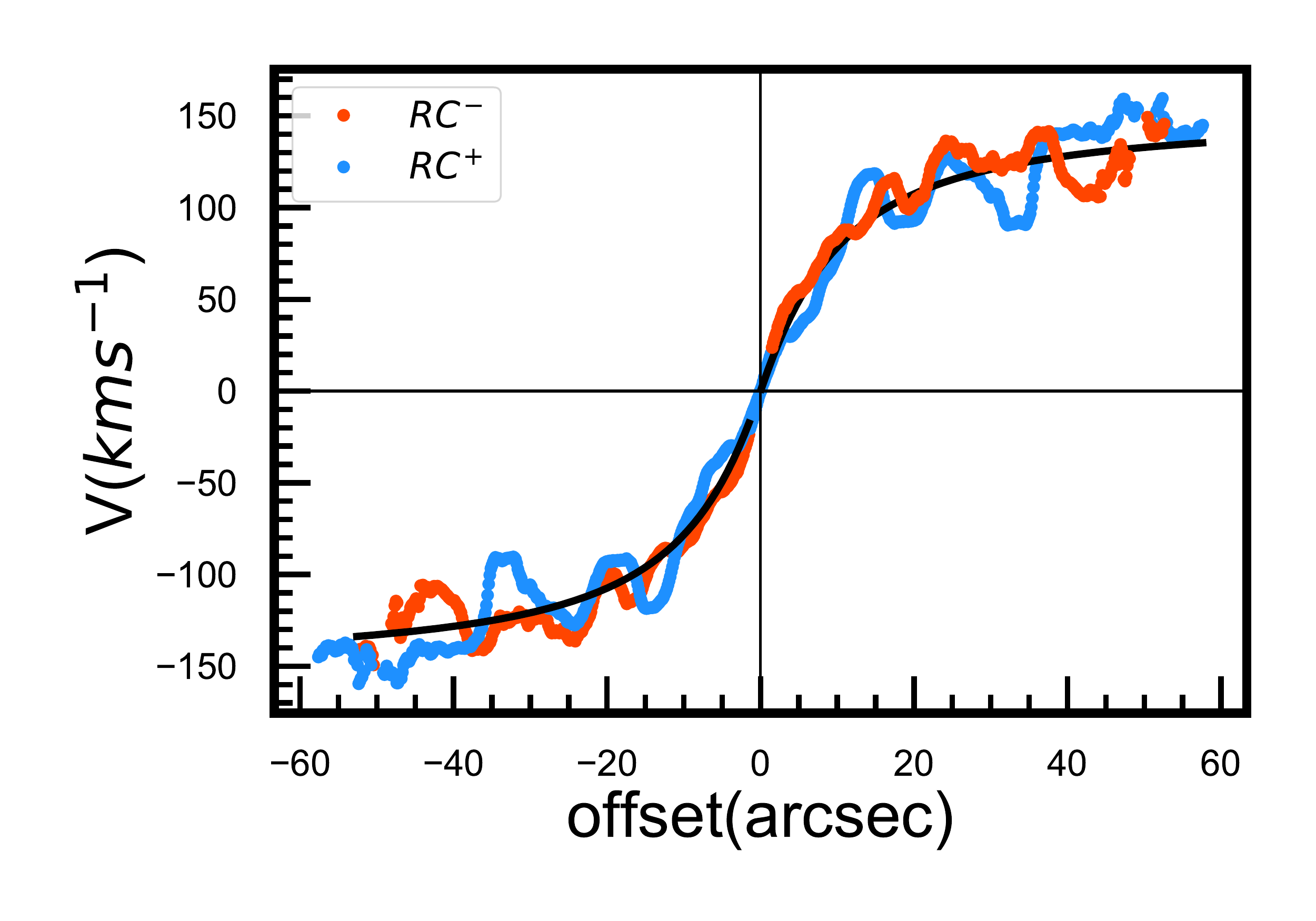}} 
\caption{The blue and red points in the above plot depict the optical rotation curve for the approaching and receding side. The points have been mirrored along 
the major axis. The smooth model optical rotation curve using equation 3 is shown using the solid black line.}	
\end{figure}
For mass modeling, we derive the hybrid rotation curve, wherein the inner region of the total rotation curve is composed of points from the $H\alpha$ rotation curve, 
and the \HI{} 21 cm data is used to define the outer points \citep{de2001high}. We have taken the optical rotation for FGC 1440 from \cite{yoachim2008kinematics}.
In order to derive a smooth curve passing through the raw optical rotation curve, we fit a simplified multi-parameter function as in  
\cite{yoachim2005kinematics,courteau1997optical},
\begin{equation}
 V(r)=V_{0} + \frac{V_{c}}{(1+x^{\gamma})^{\frac{1}{\gamma}}}
\end{equation}
where $V_{0}$ is the recession velocity of the galactic center, $V_{c}$ is the asymptotic rotation velocity (the flat part), 
$x$  is defines as $x=r_{t}/R-r_{0}$, where $r_{t}$ is the transition section between the rising and the flat part of the rotation curve and
$r_{0}$ and $\gamma$ define the center of the galaxy and degree of sharpness of transition respectively. The best-fitting parameters obtained by fitting 
Equation 3 to the raw rotation curve are detailed in Table 6. In Figure 15 we show the smooth optical rotation curve and the raw data points.

\begin{table}
%\begin{minipage}{110mm}
\caption{Best fit values describing the Multi-parameter model fitted to raw optical curve.}
\centering
\begin{tabular}{|c|c|c|c|c|}
\hline
\hline
$V^{\textcolor{red}{(a)}}_{0}$   & $V^{\textcolor{red}{(b)}}_{c}$ & $r^{\textcolor{red}{(c)}}_{0}$&$r^{\textcolor{red}{(d)}}_{t}$& $\gamma^{\textcolor{red}{(e)}}$ \\
$km s{-1}$&$\kms$&kpc &kpc&    \\
\hline
4253.5&$149.4\pm4.2$&$1.4\pm0.04 $&$12.3 \pm 0.6$&$1.3 \pm0.1$\\

\hline
\end{tabular}
\hfill{}
\label{table: table 6}
%\end{minipage}
\begin{tablenotes}
\item  $\textcolor{red}{(a)}$: Recession velocity at the galaxy center .
\item $\textcolor{red}{(b)}$:  Asymptotic rotation velocity at the flat part .
\item $\textcolor{red}{(c)}$:  Center of the galaxy. 
\item $\textcolor{red}{(d)}$:  Point of transition between the rising and the flat part.
\item $\textcolor{red}{(e)}$:  Degree of sharpness of transition.
\end{tablenotes}
\end{table}

\subsection{Dark matter models}
In this work, we parameterize the dark matter distribution using observationally motivated pseudo isothermal (PIS) 
dark matter halo  \citep{begeman1991extended, fuchs1998decomposition} dominated by a constant density core 
and, also the Navarro-Frenk-White (NFW) \cite{navarro1997universal} dark matter halo profile derived from the
cold dark matter (CDM) simulations.

\subsubsection{Pseudo-Isothermal halo model}
The density profile of the observationally motivated cored spherical pseudo-isothermal (PIS) dark matter halo is, 
\begin{equation}
 \rho(r)=\frac{\rho_{0}}{1+(\frac{R}{R_{c}})^{2}}
\end{equation}
where $\rho_{0}$ is the central density of the halo and $R_{c}$ is the core radius. The rotation curve of the PIS halo is given by

\begin{equation}
 V(R)=\sqrt{4\pi G \rho_{0} R^{2}_{c} \bigg( 1- \frac{R_{c}}{R} arctan(\frac{R}{R_{c}})\bigg)} 
\end{equation}
where $V_{\infty}=\sqrt{\bigg(4\pi G R^{2}_{c}\bigg)}$ is the asymptotic rotation velocity of the dark matter halo. The inner density distribution for PIS halo i.e $R_{c} \geq R$ 
is dominated by a constant density core .

\subsubsection{NFW halo model}
The density distribution of the cuspy NFW dark matter halo derived from the CDM simulations is 
\begin{equation}
 \frac{\rho(R)}{\rho_{crit}}=\frac{\delta_{c}}{ \bigg(\frac{R}{R_{s}}\bigg) \bigg(1 + \frac{R}{R_{s}} \bigg)^{2}   }
\end{equation}
where $\delta_{c}$ is the density of the universe at the time of the collapse of the dark matter halo,
$R_{s}$ is the characteristic scale radius, and $\rho_{crit}=3H^{2}/8\pi G$ is the critical density of the universe. 
The inner density of the NFW in the inner radii  $R\leq R_{s}$ $\rho \neq r^{-1}$, corresponding to steep and cuspy density distribution.
The rotation curve due to the NFW density distribution is,
\begin{equation}
 V(r)=V_{200} \sqrt{\frac{ln(1+cx) -cx/(1+cx)}{x[ln(1+c) -c/(1+c)]}}
\end{equation}
where, $x=R/R_{200}$, $R_{200}$ is the radius at which the mean density of the dark matter halo is 200 times the critical density. 
$V_{200}$ is the rotation velocity  at $R_{200}$. The concentration parameter is defined as $c=R_{200}/R_{s}$.

\subsection{Modified Newtonian gravity(MOND)}
 Apart from the standard dark matter model described in the above section, we use our rotation curve data to test is if just the 
 baryonic matter suffices to explain the observed rotation curve in the context of the Modified Newtonian dynamics (MOND) paradigm. 
 The net rotation curve in MOND is given by \citep{milgrom1983modification}
 \begin{equation}
  V(r)= \sqrt{\frac{1}{\sqrt{2}}( V^{2}_{gas}+  \gamma^{*}V^{2}_{*})
      \sqrt{1+\sqrt{1+ \bigg( \frac{2Ra}{ V^{2}_{gas}+  \gamma^{*}V^{2}_{stars}} \bigg) }^{2}}} 
 \end{equation}
where $a$ is acceleration and $\gamma^{*}$ is  the mass to light ratio.
\subsection{Fitting Method}

We define the likelihood function as $\rm exp(-\frac{\chi^{2}}{2})$, where $\chi^{2}$ is given by,
\begin{equation}
 \chi^{2} =\sum _{R} \frac{\bigg(V_{obs}(R) - V_{T}(R) \bigg)^{2} }{V^{2}_{err}}
\end{equation}
where $V_{obs}$ is the observed rotation curve (see top panel Figure 4), $V_{T}$ is the total rotation curve (see Equation 1) obtained by adding in quadrature the 
baryonic and the dark matter components and $V_{err}$ is the error on the observed rotation curve. For optimizing the likelihood function, we use the publicly available 
python package LMFIT \citep{newville2016lmfit}. The residuals and the corresponding reduced chi-square values $\chi^{2}_{red}$ are shown in Figures 15 and 16. 
The \HI{} rotation curve derived for FGC 1440 using 3D tilted ring modeling (section 5.2) has few points in the inner region of the galaxy, 
and the slope of the dark matter density critically depends on the shape of the rotation curve in the inner region i.e. 
a steeply rising or slowly rising rotation curve. In order to overcome this problem we use the ${H\alpha}$ rotation curve in the
inner region and the points from the \HI{} rotation curve in the outer region, where the $H\alpha$ data is not available. Although the 
shape of the $H\alpha$ rotation curve is well defined, in order to account for the scatter between 
the points, we derive a smooth representation of the $H\alpha$ rotation curve (\S 7.3). Further, to derive the hybrid rotation curve consisting of the points from the 
smooth $H\alpha$ rotation curve in the inner region and the \HI{} rotation curve in the outer region, we use B-spline method 
from python package scipy \citep{virtanen2020scipy} and create a smooth spline approximation of the data. 
Further, in order to gauge the effectiveness of using the hybrid rotation curve for estimating the dark matter parameters and mass models, we also derive 
the mass models constrained by the \HI{} rotation curve only, the results of which are shown in Appendix-B. We find that the results using only the 
\HI{} rotation curve and those using the hybrid rotation curve are comparable.
We assume uniform error bars equal to 10 \kms for the data points defining the observed rotation curve as these are typical conservative error estimates, 
see for example  \citep{de2008mass, mcgaugh2001high}, also \cite{mcgaugh2001high} show that the halo parameters are robust against precise defination of the error-bars.

In case when the dark matter distribution is parameterized using the pseudo-isothermal dark matter halo, the free parameters 
are  $\rho_{0}$ and $R_{c}$, and mass to light ratio $\gamma^{*}$ along with $\rho_{0}$ and $R_{c}$ 
in case of mass, models derived keeping the mass to light ratio as a free parameter. When the dark matter density is parameterized 
using the NFW halo, the free parameters are  $c$ and $R_{200}$, and  $\gamma^{*}$ is the free parameter along with $c$ and $R200$ in
case of mass models in which the mass to light ratio is kept as
a free parameter.

\subsection{Results from mass modeling}
This section presents the mass models constructed using the SDSS optical g-band and NIR K-band rotation curve in conjugation with the derived $\rm \HI$ and 
the total rotation curve as detailed in the above sections. For each of the photometric band, we fit both the NFW and PIS dark matter halo profile and discuss the 
following cases:

\begin{itemize}
 \item \textbf{The constant $\rm \gamma^{*}$} We derive $\gamma^{*}$ using population synthesis models as described by \cite{bell2001stellar}.\\ 
 a)$\rm diet-Salpeter$ $\rm IMF$  which gives highest disc mass for a given photometric band \citep{de2008high} and\\ 
 b)$\rm Kroupa $ $\rm IMF$ which produces lower disc mass \cite{kroupa2001variation} as compared to $'diet-SalpeterIMF'$.
 \item \textbf{Free $\gamma^{*}$} In this model, the $\gamma^{*}$ is kept as a free parameter along with the parameters corresponding to the dark matter density profile.
 \item \textbf{Maximum Disc} We scale the stellar rotation curve by scaling the $\gamma^{*}$ such that the observed rotation curve in the inner region is entirely due to the stellar component. The maximum disc model sets the lower limits on the dark matter density.
 \item \textbf{Minimum Disc} We set the contribution of the gaseous disc and stars to zero and attribute the observed rotation curve to be entirely due to the
 underlying dark matter distribution. The minimum disc model sets the upper limit on the dark matter density.
\end{itemize}

\begin{table*}
\hspace{-65.5mm}
\begin{minipage}{110mm}
\hfill{}
\caption{Dark matter density parameters derived from mass-modeling using the optical g-band and NIR K band photometry using the hybrid (\HI{} + H$\alpha$) rotation curve.}
\centering
\begin{tabular}{|l|c|c|c|c|c|c|c|c|c|c|}
\hline
\hline
Model             & $c^{\textcolor{red}{(a)}}$           &$R^{\textcolor{red}{(b)}}_{200}$         & $\gamma^{*\textcolor{red}{(c)}}$ &$\frac{V_{max}}{V_{200}}^{\textcolor{red}{(d)}}$     & $\chi^{2\textcolor{red}{(e)}}_{red}$     & $\rho^{\textcolor{red}{(f)}}_{0}\times10^{-3}$  &$R^{\textcolor{red}{(g)}}_{c}$ & $\gamma^{*\textcolor{red}{(h)}}$& $\frac{R_{c}}{R_{d}}^{\textcolor{red}{(i)}}$     &$\chi^{2\textcolor{red}{(j)}}_{red}$ \\
                  &             &(kpc)             &             &              &                  &  $M_{\odot}/pc^{3}$      &(kpc)   &           &      &                                  \\
g-band            & NFW  profile&                  &             &              &                &PIS profile               &        &             &      & \\
\hline

$'diet'$ Salpeter	&$5.06\pm 0.5	$&$	84.1\pm0.3$	&$	3.8	$&$	2.3$&$	0.1	$& $	56.6\pm0.4	$ &$	2.3\pm0.01	$ &$	3.8	$&$	0.5$&	$0.1$	\\
Kroupa IMF 	&$	5.5 \pm0.06	$& $84.4\pm0.3$	& $2.7$ & $2.3$ & $0.1$ & $	66.1\pm0.6$  & $2.2\pm0.01$ &$	2.7	$&$	0.5	$&$	0.01$	\\
Free $\gamma^{*}$ 	&$3.8\pm0.1	$&$	85.0\pm0.3$	&$	6.4\pm0.2	$&$	2.3	$&$	0.05	$&$	58.0\pm0.61	$&$	2.3\pm0.01	$&$	3.6\pm0.05	$&$	0.5	$&$	0.01$	\\
Maximum Disc 	&$	0.06\pm0.04	$&$	218.4\pm9.6$	&$	14	$&$	0.07	$&$	0.3	$&$	1.4\pm0.05	$&$	19.2\pm0.8	$&$	14	$&$	4.3	$&$	0.2$	\\
Minimum Disc	&$	5.7\pm 0.06	$&$	96.1\pm0.3$	&$	0	$&$	2.0	$&$	0.1	$&$	71.9\pm0.43	$&$	2.4\pm0.0	$&$	0	$&$	0.5	$&$	0.9$	\\
\hline
K-Band          &                    &                  &        &           &                      &   &\\
\hline
$'diet'$ Salpeter	&$2.8\pm0.09	$&$	97.0\pm1.3$	&$	0.5	$&$	2.01	$&$	0.4	$&$	13.9\pm0.4	$&$	4.9\pm0.09	$&$	0.85	$&$	1.9$&$	0.14$	\\
Kroupa IMF 	&$	3.8\pm0.09	$&$	91.6\pm0.8$	&$	0.6	$&$	2.1	$&$	0.3	$&$	24.3\pm0.5	$&$	3.7\pm0.05	$&$	0.6	$&$	1.	$&$	0.08$	\\
Free $\gamma^{*}$ 	&$5.6\pm0.2	$&$	87.9\pm0.6$	&$	0.2\pm0.04	$&$	2.2	$&$	0.2	$&$	38.8\pm2.7	$&$	3.02\pm0.11	$&$	0.4\pm0.03	$&$	1.2	$&$	0.1$	\\
Maximum Disc 	&$	1.6\pm0.10	$&$	128.9\pm3.6	$&$	1.1	$&$	1.5	$&$	0.6	$&$	8.5\pm0.3	$&$	6.7\pm0.2	$&$	1.1	$&$	2.6	$&$	0.3$	\\
Minimum Disc	&$	5.7\pm0.06	$&$	96.1\pm0.3$	&$	0	$&$	2.0	$&$	0.1	$&$	71.8\pm0.4	$&$	2.4\pm0.0	$&$	0	$&$	0.9	$&$	0.1$\\
\hline
MOND                       &                                    &                                        &                &     &    &    &    \\
\hline
        &$a^{\textcolor{red}{(k)}}$ &  $\gamma^{*\textcolor{red}{(l)}}$  &  $\chi^{2\textcolor{red}{(m)}}_{red}$  &                &     &    &    & \\
        &$ms^{-2}$  &                                    &                                        &                &     &    &    &\\        
\hline
g-Band              & $0.42 \times 10^{-10}$         & $12.7 \pm 0.1 $&$0.2$  &            &    &    &    &    \\    
$g-Band^{a=fixed}$  &$1.2 \times 10^{-10}$                       & $4.1 \pm 0.1$      & 1.8        &    &    &    &    \\
K-Band              &$0.85 \times 10^{-10}$         &1.0                  & 0.5       &    &    &    &     \\             
$K-Band^{a=fixed}$  &$1.2 \times 10^{-10}$ & $0.6 \pm 0.01$    &0.8&                 &            &    &    &    &      \\
\hline
\end{tabular}  
\hfill{}
\label{table: table 7}
\end{minipage}
\begin{tablenotes}
\item  $\textcolor{red}{(a)}$: Concentration parameter of the NFW profile\\
\item $\textcolor{red}{(b)}$:  Radius at which the mean density equal to 200 times the critical density.\\
\item $\textcolor{red}{(c)}$:  Mass to light ratio derived using population synthesis models or estimated as a free parameter. \\
\item $\textcolor{red}{(d)}$:  Ratio of the asymptotic velocity to the  velocity at $\frac{V_{200}}{\kms}=0.73\frac{R_{200}}{kpc}$\citep{navarro1997universal}\\
\item $\textcolor{red}{(e)}$:  Reduced chi-square value corresponding to the fit. \\
\item $\textcolor{red}{(f)}$:  The central dark matter density of the PIS dark matter halo model\\
\item $\textcolor{red}{(g)}$:  The core radius of the PIS dark matter halo model\\
\item $\textcolor{red}{(h)}$:  Mass to light ratio derived using population synthesis models or estimated as a free parameter. \\
\item $\textcolor{red}{(i)}$:  Ratio of the core radius and the disc scalelenght.\\
\item $\textcolor{red}{(j)}$:  Reduced chi square value corresponding to the fit.\\
\item $\textcolor{red}{(k)}$:  Acceleration per lenght in MOND.\\
\item $\textcolor{red}{(l)}$:  Estimated Mass to light ratio in MOND.\\
\item $\textcolor{red}{(m)}$:  Reduced chi square corresponding to the fit.\\
\end{tablenotes}
\end{table*}

\begin{figure*}
\resizebox{190mm}{180mm}{\includegraphics{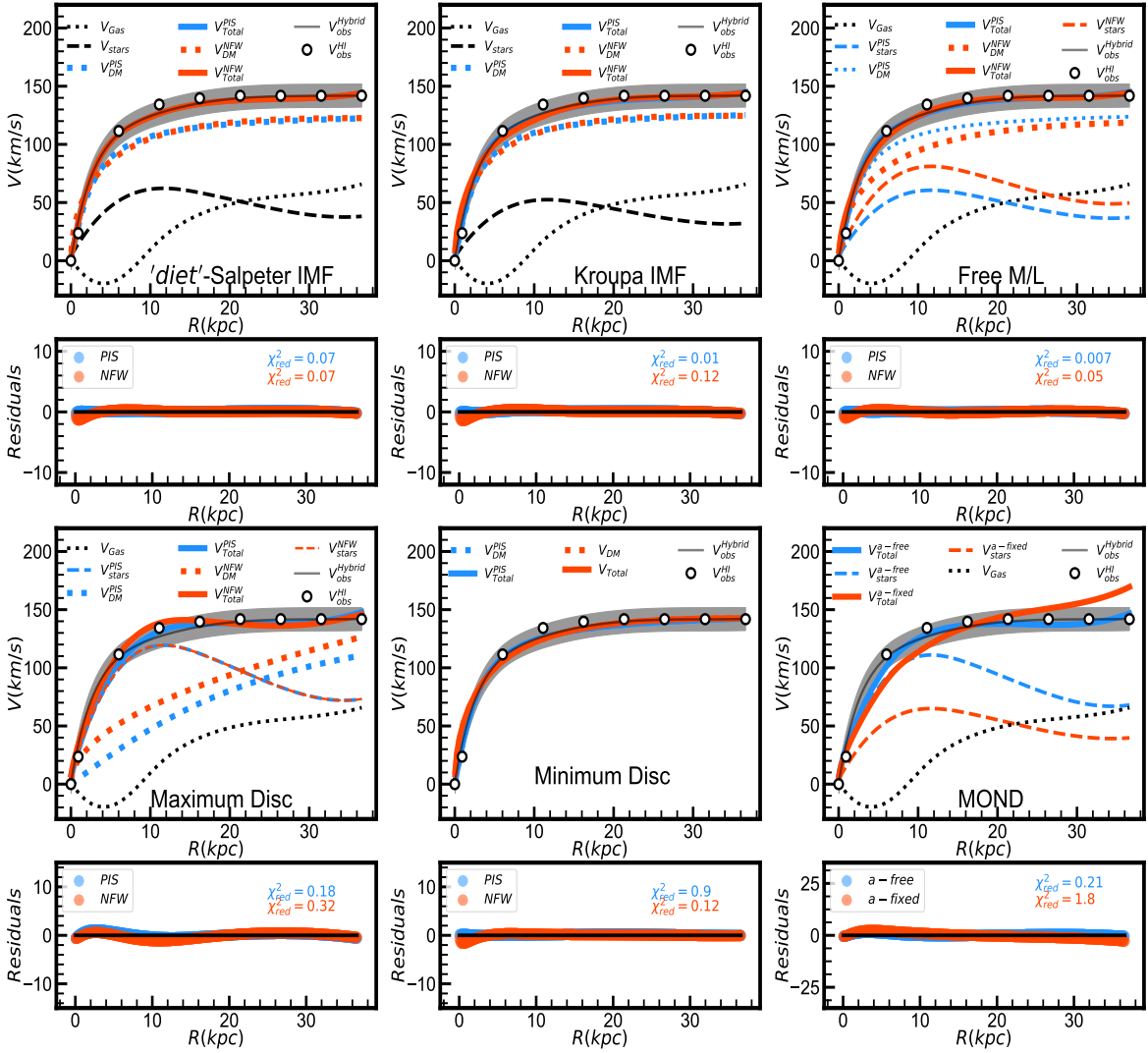}} 
\caption{We present the mass model of the galaxy FGC 1440 derived using SDSS g-band photometry. The mass models are constrained using the
hybrid rotation curve.}

\end{figure*}
\begin{figure*}
\resizebox{190mm}{180mm}{\includegraphics{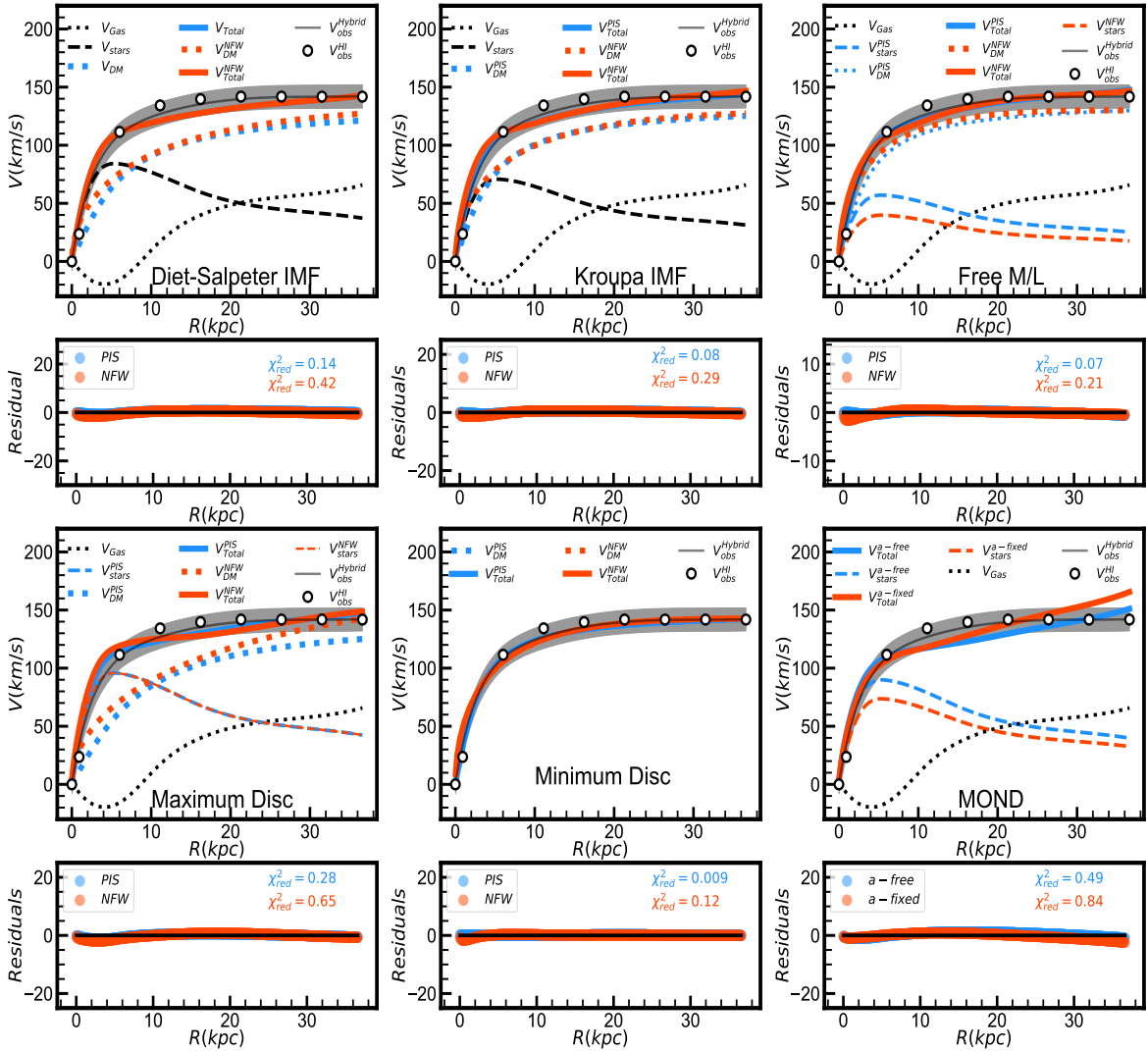}} 
\caption{We present the mass-model of the galaxy FGC 1440 derived using UKIDSS K-band photometry. The mass models are constrained using the hybrid rotation 
curve.}
\end{figure*}

In Figure 16, we present mass models constructed using the stellar rotation curve derived using g-band Photometry and the \HI{} rotation curve. We find that the reduced chi square$(\chi^{2}_{red})$ for the cored pseudo-isothermal dark matter halo is lower than the cuspy NFW dark matter halo, only in the case of the model derived using $'diet'-Salpeter$  IMF both the halos give similar $(\chi^{2}_{red})$ values. For mass models using PIS halo, we find that the Kroupa IMF gives lower reduced $(\chi^{2}_{red})$ as compared to the 
$'diet'-Salpeter$ IMF, indicating that the mass distribution is possibly dominated by dark matter.
In mass models for PIS halo with $\gamma^{*}$ as a free parameter, $\gamma^{*}$ values tend to values closer to that derived using the $'diet'-Salpeter$ IMF. in case of the mass models with NFW halo and  
$\gamma^{*}$ kept as a free parameter, we find that the $\gamma^{*}$ tends to values higher than that derived using the stellar population synthesis models.
In the case of the maximum disc models, we have scaled the stellar rotation curve by a factor of 14 to maximize the contribution of the stellar disc and set the baryonic to zero in the case of the minimum disc models.\\

Similarly, in Figure 17, we present the results from mass models using the stellar rotation curve in K-band. The $\chi^{2}_{red}$ for the mass model with PIS dark matter halo is systematically lower than that corresponding to the NFW dark matter halo. The $\chi^{2}_{red}$ for the Kroupa IMF is lower than that for $'diet'-Salpeter$ IMF. Whereas in the case of mass models with $\gamma^{*}$ as a free parameter, $\gamma^{*}$ is even lower than that derived using 
Kroupa IMF in the case of both PIS halo and NFW halo, possibly indicating the mass models prefer lighter IMF in K-band. We scale the stellar rotation curve by a 
a factor of 1.1 to derive the maximum disc model and set the baryonic contribution to zero for the minimum disc models. 

The definition of compact dark matter halo follows from the ratio of the core radius to the disc scalelength $R_{c}/R_{d}<2$ 
\citep{banerjee2017mass}. In table  7, we have indicated 
$R_{c}/R_{d}$ for each model, in all the cases other than the maximum disc case in K-band $R_{c}/R_{d}$is less than 2. 
We also note that the definition of a compact halo is not independent of the choice of IMF, as models which prefer maximum disc in a given photometric
band have a larger core radius and less compact dark matter halo.\\

For models with constant IMF, we find that the concentration parameter ranges between 2.8 (diet-Salpeter K-band) and 
5.5 (Kroupa IMF in g-band). Similar to the compactness parameter, the IMF models preferring higher disc masses have smaller concentration parameters. 
The scaling between asymptotic rotation velocity $V_{max}$ and concentration parameter given in \cite{bottema2015distribution} 
$c_{exp}=55.5(V_{max}/[km s^{-1}])^{-0.2933}$, gives us $c=12.97$. We note that the value of concentration parameters derived in this work 
are much lower than the value predicted by the scaling.Similarly the using the scaling between  
$V_{max}$ and $R_{200}$, where $R_{200}=0.0127(V_{max})^{1.37} c_{exp}$ gives $R_{200}=146kpc$ 
which is closer to the values of $R_{200}$ obtained in case of maximum disc models. 
We also derive mass models using \HI{} rotation curve derived using the tilted ring models, see Appendix- B, we find that the reduced 
$\chi^{2}_{reduced}$ for the PIS halo is smaller than that obtained for the NFW halo.

We further compare our results with theoretical predictions between the ratio of $V_{max}/V_{200}$ and concentration parameter 
from \cite{dutton2014cold}. The relation between $V_{max}/V_{200}$ and concentration parameter $c$ is defined as 
$V_{max}/V_{200}= 0.216c_{200}/f(c_{200})$, where $f(c_{200})= ln(1+c) - c/(1+c)$. Using the the values of concentration parameter 
from Table. 7, we find that the value of $V_{max}/V_{200}$ is close to 1.

We test for the correlation between the dark matter core radius and the disc scalelenght using the relation given in \cite{donato2004cores} 
$log(R_{c})=(1.05 \pm 0.11) log(R_{d}) + (0.33 \pm 0.04)$. With K-band disc scalelenght $R_{d}=2.58 kpc$ we get a core radius equal to 5.78 kpc, which is closer to the 
diet-Salpeter and the maximum disc case. Similarly the g-band scalelenght gives core radius equal to 10.72 kpc.

We compare the parameters $V_{\infty}=\sqrt{4 \pi G \rho_{0}R^{2}_{c}}$ for the PIS halo and $R_{s}=R_{200}/c$ of FGC 1440 with the that of other 
superthins in the literature as $R_{s}$ and $V_{\infty}$ constitute single parameter encompassing both the fit parameters.
In the study of three superthins \cite{banerjee2017mass} find $V_{\infty}$ equal to 110\kms , 112 \kms and 99 \kms for 
UGC7321, IC5249 and IC2233 respectively. In another another study,  \cite{kurapati2018mass} find $V_{\infty}$ equal to 
82.7 \kms for FGC 1540. We find $V_{\infty}=135$ for FGC 1440. \cite{banerjee2017mass} find $R_{s}$ equal to 
8.55 and 22.6 for UGC 7321 and IC 5249, \cite{kurapati2018mass} find $R_{s}$ equal to 5.25 for FGC 1540. For FGC 1440, we find $R_{s}$ equal to  24.3.

\subsubsection{Mass models MOND}
In the last panels of figures 16 and 17, we have shown the mass models derived using MOND. Keeping both $a$ and $\gamma^{*}$ as a free parameter we find that the,
acceleration parameter $a=0.42 \times 10^{-10}ms^{-2}$ and $\gamma^{*}=12.7$ in g-band. Similarly in K-band the acceleration parameter $a=0.85 \times 10^{-10} ms^{-2}$ and
$\gamma^{*}=0.97$. We note that in the case when both the $a$ and $\gamma^{*}$ are kept as free parameters, $\gamma^{*}$ tends to values maximizing the disc mass, i.e., 
they are closer to the maximum disc case of dark matter models. We also try the case in which we fix the value of $a=1.2 \times 10^{-10}ms^{-2}$ and only vary $\gamma^{*}$, 
we find that the value of $\gamma^{*}$ is equal to 4.14 and 0.65 in g-band and K-band respectively, and is closer to the values derived using population synthesis models.

\section{Vertical structure of FGC 1440.}
\begin{figure*}
\hspace{-13.5mm}
\resizebox{190mm}{70mm}{\includegraphics{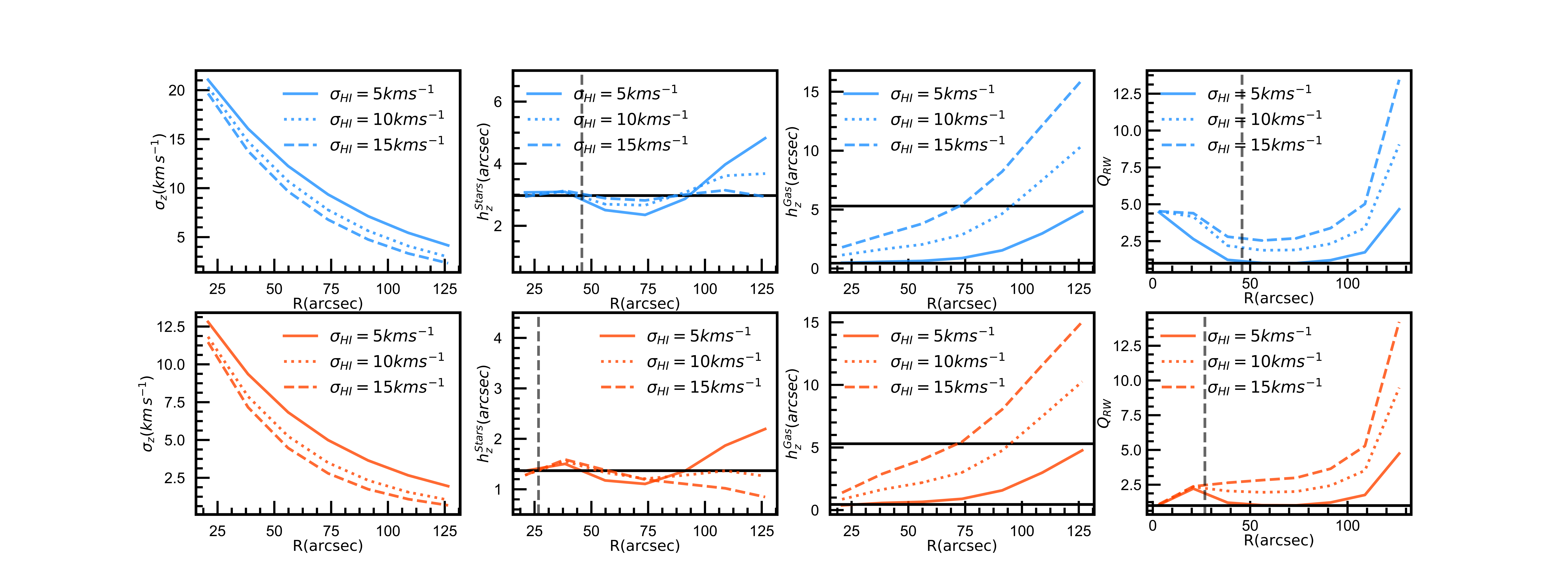}} 
\caption{ The plots show the vertical velocity dispersion $\sigma_{z}$, the modeled stellar $(h^{Stars}_{z})$ and \HI{} scaleheights $(h^{Gas}_{z})$, and 
 the stability parameter $(Q)$. The top panel in blue ink depicts the results for g-band and lower panel in red, the results in K-band. The vertical dashed line marks the 3$R_{d}$
 in g-band and K-band respectively. The horizonal black line in panel depicting $h^{Stars}_{z}$ marks the observed stellar scaleheight. The horizonal line in plot showing 
 $h^{Gas}_{z}$ marks the upper and the lower limit on the \HI{} scaleheight derived using the tilted ring modeling.} 
\end{figure*}
We model the galaxy disc as a co-planar, co-axial gravitationally coupled $star + gas$ system under the influence of the force field of external dark matter halo.  
Solving the two-component Jeans equation using the methods outlined in \cite{10.1093/mnras/stab155} and \cite{komanduri2020dynamical} we derive the stellar vertical velocity dispersion $(\sigma_{z})$  as function of 
radius constrained by the observed stellar scaleheight. We use the best fitting mass model, i.e., the pseudo-isothermal (PIS) profile with stellar surface density scaled by Kroupa IMF, along with the observed \HI{} surface density and the \HI{} dispersion as the input parameters. From the 3-D models of the \HI{}
data cube, we find that the \HI{} dispersion is constrained in the range  $5 \kms <\sigma_{HI} <15 \kms$, so we model the stellar vertical dispersion fixing the \HI{}
dispersion at 5\kms{}, 10\kms, 15\kms at all radius. The stellar dispersion is modeled as an exponential function $\sigma_{z}(R)=\sigma_{0}e^{\frac{-R}{\alpha R_{d}}}$, 
where $\sigma_{0}$ is the central dispersion and the $\alpha$ is the steepness parameter. The values of the $\sigma_{0}$ and $\alpha$ derived using the mass models in
g-band, and K-band are summarized in Table 8. We find that the central value of the stellar dispersion is not sensitive to the variation of the $\sigma_{HI}$ in the allowed range 
, but the steepness parameter is. The steepness parameter $(\alpha)$ varies between for $3.2\,-\, 4.2$ in g-band and  $4.2\,-\, 6.3$ in K-band for $5\kms\sigma_{HI}<15\kms$.
For the different values of the $\sigma_{HI}$, we find that the modeled stellar scaleheight agrees with the observed scaleheight up to $3R_{d}$. For the values of
$\sigma_{HI}=5 \kms$ we find that that the the model predicted \HI{} scaleheight (panel 3 in Figure 17) is
constrained between the limits obtained from tilted ring modeling $(0.45\farcs<h_{z}<5.3\farcs)$.

\begin{table}
\begin{minipage}{110mm}
\hfill{}
\caption{Values of vertical stellar dispersion.}
%\centering
\begin{tabular}{|l|c|c|c|c|}
\hline
\hline
Parameter                  &           g-Band       &             & K-Band         &            \\
\hline
                           &         $\sigma_{0}$   &  $\alpha$   &  $\sigma_{0}$  &  $\alpha$  \\  
			   &          \kms          &             &   \kms         &            \\
\hline    
$\sigma_{HI}$=5  \kms      &          29.0          &4.2         &18.6            &6.3   \\   
$\sigma_{HI}$=10 \kms      &          29.7          &3.6          &19.1            &4.8  \\
$\sigma_{HI}$=15 \kms      &          29.9          &3.2         &20.1            &4.2   \\ 
\hline
\end{tabular}
\hfill{}
\label{table: table 8}
\end{minipage}
\end{table}

\section{Discussion}

\begin{itemize}

 \item In this Section we discuss the main dynamical properties of FGC 1440, and compare them with those determined in the 
       literature for other superthin galaxies.

\begin{itemize}

\item \textbf{Disc Heating} \\
It has been shown that bars, spiral arms, and globular clusters play an important role in disc heating. Bars and spiral arms heat the disc in the radial direction whereas globular clusters heat the disc isotropically in both the radial and vertical direction \citep{aumer2016age, jenkins1990spiral, grand2016spiral, saha2014disc}. Further, 
\cite{banerjee2013some} showed that a compact dark matter halo regulates the distribution of the stars in the vertical direction and gives rise to superthin 
disc structure. \cite{10.1093/mnras/stab155}  show for a sample of superthin galaxies that the absolute values of the vertical stellar velocity
dispersion is small as compared to the Milky way. And upon comparing the the ratio of vertical velocity dispersion to 
the total rotation velocity ($\frac{V_{Rot}}{\sigma_{z}}$), they find that the value of $\frac{V_{Rot}}{\sigma_{z}}$ is comparable to stars in the thin disc of Milky way, 
indicating that superthin galaxies are dynamically cold systems. Thus, it is imperative to compare the value of $\frac{V_{Rot}}{\sigma_{z}}$ for FGC 1440 with previously
studied superthin galaxies to quantify the effect of the disc heating.
Using the
values of vertical velocity dispersion constrained using optical photometry; we find 
that the ratio $\frac{V_{Rot}}{\sigma_{z}}$ for FGC 1440 is comparable to the sample of superthin galaxies studied in \cite{10.1093/mnras/stab155} equal to 5.0, except for  
UGC 7321 which has $\frac{V_{Rot}}{\sigma_{z}}=10$. Similarly using the values of the velocity dispersion constrained with the g-band and K-band data we find that value 
of $\frac{V_{Rot}}{\sigma_{z}}=8$ for FGC 1440 compared to 10 for other superthin galaxies, except for IC2233 which has $\frac{V_{Rot}}{\sigma_{z}}=14$ and UGC00711 which 
has $\frac{V_{Rot}}{\sigma_{z}}=5$. 

 \item \textbf{Disc dynamical stability}\\ 
The disc dynamical stability $Q$ as quantified by \cite{toomre1964gravitational}  $Q= \frac{\kappa \sigma  }{\pi G \Sigma}$, where Toomre stability criterion $'Q'$  
is a subtle balance between the epicyclic frequency of the self-gravitating matter $\kappa$, the radial velocity dispersion $\sigma$ and the 
surface density $\Sigma$. The epicyclic frequency $\kappa$ at a radius R is defined as $\kappa^2(R)= \big( R\frac{d\Omega (R)^{2}}{dR} + 4\Omega (R)^{2}\big)$, where $\Omega$
is the angular frequency defined as $\Omega (R)^{2}=\frac{1}{R}\frac{d\Phi_{Total}}{dR}= \frac{V^{2}_{rot}}{R^{2}}.$ $\Phi_{Total}$ is the total gravitational 
potential and $V_{Rot}$ is the total rotation velocity. 

\cite{garg2017origin} show that dark matter plays a decisive role in regulating the stability of the low surface brightness 
galaxies and that in the absence of dark matter, the galaxy disc would be susceptible to axis-symmetric instabilities. Further, in 
\cite{10.1093/mnras/stab155} it has been shown for a sample of superthin galaxies that the median values of stability for the $star+gas$ system is higher than the
typical spiral galaxies studied by \cite{romeo2017drives}. This section compares the value of the stability of the $star+gas$ disc of FGC 1440 with that of previously
studied superthin galaxies. Using the two-component stability parameter derived in \cite{romeo2011effective}, we compute the dyanmical stability of the star+gas system for FGC 1440.

The two-component disc stability parameter $Q_{RW}$ appraising the stability of the composite $star+ gas$ disc is given by
\begin{equation}
\frac{1}{Q_{RW}} = \left\{
                \begin{array}{ll} \frac{ W_{\sigma} }{T_{s}Q_{s}} + \frac{1}{T_{g}Q_{g}}  \hspace*{0.5cm} if \hspace*{0.5cm}  T_{s}Q_{s} > T_{g}Q_{g}\\  
         \frac{1}{T_{s}Q_{s}} +\frac{W_{\sigma}}{T_{s}Q_{s}}                            \hspace*{0.5cm} if \hspace*{0.5cm} T_{s}Q_{s} < T_{g}Q_{g}  
         \end{array}
              \right.
\end{equation}
\noindent where the weight function $W$ is given by
\begin{equation}
 W_{\sigma} =\frac{2\sigma_{s} \sigma_{g}}{\sigma_{s} ^{2} + \sigma_{g}^ {2}}
\end{equation}
The thickness correction is defined as;
\begin{equation}
 T \approx 0.8 + 0.7 \frac{\sigma_{z}}{\sigma_{R}}
\end{equation}

In the above equation $\sigma_{s}$ and $\sigma_{g}$ are the velocity dispersion of the stars and gas respectively. $Q_{s}$ and $Q_{g}$ are the Toomre Q of the 
stellar disc and gas disc respectively . A value of $Q_{RW} >1$ indicates that the composite $stars+gas$ disc is stable against axi-symmetric perturbations.

We calculate the disc dynamical stability parameter $(Q)$ (panel 4 in Figure 18) against 
local, axis-symmetric perturbations of the galaxy as a function of the radius using the two-component stability parameter derived in \cite{romeo2011effective}.
We compute the stability parameter using value of \HI{} dispersion; $\sigma_{\HI{}}=5 \kms \,10 \kms \, 15 \kms $ in both g-band and K-band.

We find that the minimum value of 
$Q$ in g-band is 1.0, 1.9, 2.5 corresponding to $\sigma_{\HI{}}=5 \kms\,10 \kms\, 15\kms$. Similarly, in K-band the minimum value of $Q$ is equal to 1.1 
$\sigma_{\HI{}}=5 \kms\,10 \kms \, 15\kms$. But we note that in g-band $Q > 2$ for $R<3R_{d}$, indicating that the stellar disc is possibly stable, 
whereas in K-band the disc is closer to marinal stability.  We also note that in both g-band and K-band galaxy disc is closer to marginal stability for lower 
values of the \HI{} dispersion $\sigma_{HI}=5\kms$ than for $\sigma_{HI}=10\kms$ or $\sigma_{HI}=15\kms$ for $R<3R_{d}$.
Further, we find that the the value of stability parameter $Q(R<3R_{d})$ is lower than the median value of 5.5 \citep{10.1093/mnras/stab155}for previously studied sample 
of superthin galaxies and is closer to the median value of Q for calculated for a sample of spiral galaxies equal to $2.2 \pm 0.6$  by \citep{romeo2017drives}.

\begin{figure*}
%\hspace{-13.5mm}
\resizebox{180mm}{55mm}{\includegraphics{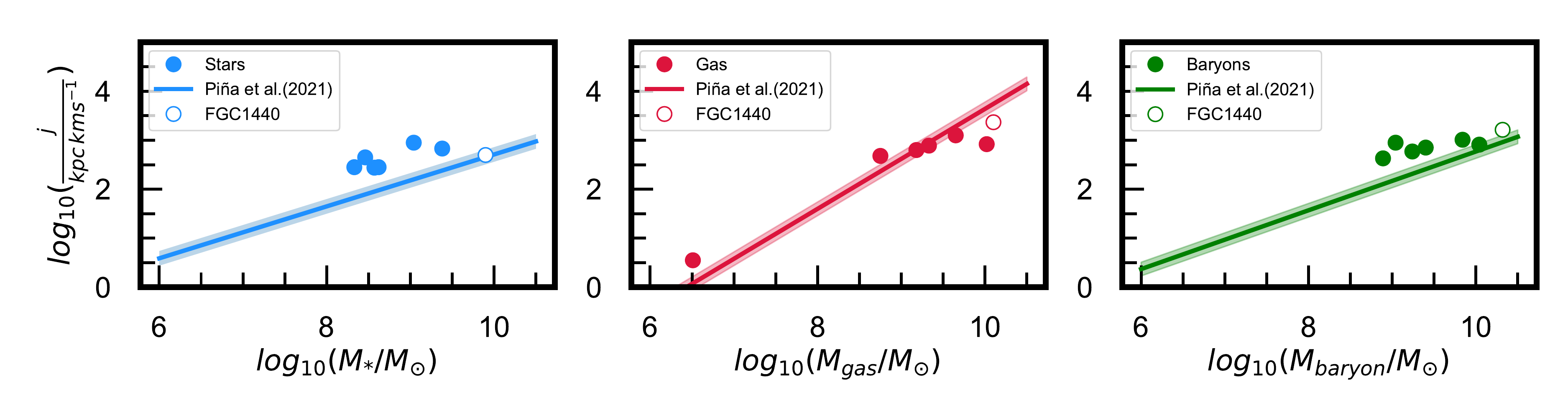}} 
\caption{The plots indicate the Fall relation for superthin galaxies (filled-points) and compare it with that of FGC 1440 (open-point). 
The straight line shows the best-fitting Fall relation and the shaded region indicates the intrinsic scatter obtained by \protect \cite{pina2021baryonic}  for disc galaxies.} 
\end{figure*}

\item \textbf{Specific angular momentum}\\ 

The Fall relation \citep{fall1980formation, romanowsky2012angular, posti2018angular} connects the mass to the specific 
angular momentum in the disc galaxies. The Fall relation is well established observationally for disc galaxies of diverse morphologies
\citep{posti2019galaxy, pina2021baryonic, marasco2019angular}, also see \cite{kurapati2018angular}. The study consisting of sample of superthin galaxies and low surface brightness 
galaxies \cite{jadhav2019specific} showed that the specific angular momentum of the low surface brightness galaxies is higher 
than ordinary disc galaxies indicating that possibly high specific angular momentum drives the superthin disc stucture. We calculate  $j-M$ values
for FGC 1440 to see how the specific angular momentum of extremely thin disc galaxy FGC 1440 compares with a sample of previously studied 
superthin galaxies  \citep{jadhav2019specific} and a larger sample of disc galaxies \citep{pina2021baryonic}. 
Given a rotation curve $V(R)$ and a surface density profile $\Sigma(R)$, the specific angular momentum within a radius R is given by

\begin{equation}
 j_{i}(<R)= \frac{2 \pi \int^{R} _{0} R^{'2} \Sigma_{i}(R^{'}) V_{i}(R^{'})dR^{'} }{ 2 \pi \int^{R} _{0} R^{'} \Sigma_{i}(R^{'}) dR^{'}}
\end{equation}

In the above equation $i$ indexs over stars$(*)$, gas $(g)$ and baryons ($b$, defined as the sum of stars and gas). We use the K-band photometery 
from Table. 5 for computing the $j_{*}$ and $M_{*}$, and \HI{} surface density in Figure. 4 to compute $j_{g}$ and $M_{g}$. We find that 

$log_{10}\frac{j_{*}}{kpc \, \kms}$ is equal to 2.7  for the 
stellar mass equal to $7.9 \times 10^{9} M_{\odot}$ and the specific angular momentum of the gas disc 
$log_{10}\frac{j_{g}}{kpc \, \kms}$ is equal to 3.4  and mass of 
the gas disc is  $1.3 \times 10^{10} M_{\odot}$. In Figure. 19 we compare the Fall relation 
for FGC 1440 $(open-points)$ with other superthin galaxies $(filled-points)$, the data points are taken from \cite{jadhav2019specific}.
We also plot the $j-M$ relation $(straight-line)$  obtained for stars, gas and baryons by \cite{pina2021baryonic} for a large sample of disc 
galaxies.

We find that unlike other superthin galaxies, which are outliers in the $log_{10}(j_{*}) - log_{10}(M_{*})$ relation for the ordinary disc galaxies, 
FGC 1440 closely follows the regression line obtained by \cite{pina2021baryonic} in the $log_{10}(j_{*}) - log_{10}(M_{*})$,
$log_{10}(j_{b}) - log_{10} (M_{b})$ and the $log_{10}(j_{g}) - log_{10}(M_{g})$ relation.

\end{itemize}

\end{itemize}

\section{Conclusion}
We have presented the analysis and results pertaining to $\rm \HI$ imaging of ultra-flat galaxy FGC 1440 observed using GMRT. 
\begin{itemize}
\item We fit the busy function to the HI spectrum and find that the velocity widths 20$\%$ and the 50$\%$ of the peak maximum are 295 \kms and 306 \kms, respectively. We find that the total flux density is 10.46 $Jy \kms$.
\item From our preliminary analysis of the HI data cube, we find that the moment 0 and moment 1 maps show that the $\rm \HI$ disc is slightly warped on the north-eastern side.
\item We use our final $\rm \HI$ data to construct tilted rings model of the \HI{} emission and derive the kinematic parameters using 
TiRiFic \citep{jozsa2012tirific} and FAT \citep{kamphuis2015fat}. We find that a model with radially varying inclination equal to 
90$^{\circ}$ in the inner rings and 85$^{\circ}$ for the outer rings gives a better description of the data. 
The best fit value of the position angle is 53.6$^{\circ}$. We find that FGC 1440 has a slowly rising rotation curve with an asymptotic rotation velocity equal to 141.8 \kms.

 \item By manually comparing the PV diagrams at different offsets, we find that the HI velocity dispersion lies in between $5\kms <\sigma < 15 \kms$ and the measurement of the scaleheight is limited by the resolution of the synthesized beam, $h_{z}< 5.3\farcs$.
 \item By comparing the data with models with a varying inclination and scaleheight, we find that FGC 1440 possibly hosts a thin \HI{} disc warped along the line of sight. \item We use the total rotation velocity along with the stellar photometry to derive the mass models in optical g-band and NIR 
UKIDSS K-band.  We find that the mass models derived using cored pseudo-isothermal dark matter halo in conjugation with stellar rotation curves derived using Kroupa initial mass function give better fits to the observed rotation curve.
\item We also derive mass models in modified Newtonian paradigm, we find that models in which both acceleration and the $\gamma^{*}$ are kept as free parameters, the values of acceleration are lower than $1.2 \times 10^{-10}\,ms^{-2}$ and the mass to light ratios tend to values maximizing the disc mass. 
Whereas as models in which a is fixed at $1.2 \times 10^{-10}\,ms^{-2}$, $\gamma^{*}$ tends to values predicted by stellar population synthesis models.

\item Using the observed stellar scaleheight as we constrain the vertical velocity dispersion in g-band and K-band. We find that the value of 
central dispersion is equal to 29.0 \kms in g-Band and 18.6 \kms in K-band. We note the values of vertical dispersion are comparable to the values of dispersion quoted in 
\cite{10.1093/mnras/stab155}. 

\item Using the two-component stability parameter proposed by \cite{romeo2011effective}, we calculate the stability factor for FGC 1440, we find that Q-value is greater than 1 
for $R<3R_{d}$, indicating that the galaxy is stable against axis-symmetric instabilities. 
The value of Q for FGC 1440 is lower than the median value of Q for superthin 
galaxies, equal 5.5 \citep{10.1093/mnras/stab155} and is closer to the median value of Q for calculated for a 
sample of spiral galaxies equal to $2.2 \pm 0.6$  by \citep{romeo2017drives}.  We find that that in spite of the large axial ratios the values of 
$\frac{V_{Rot}}{\sigma_{z}}=5$ in g-band and $\frac{V_{Rot}}{\sigma_{z}}=8$ in K-band are comparable to 
the superthin galaxies previously studied in the literature. 

\item Inspecting the Fall relation, for ordinary disc galaxies, we find that FGC 1440 follows the 
regression line for the $log_{10}(j_{*})- log_{10}(M_{*})$, $log_{10}(j_{b}) - log_{10}(M_{b})$ and $log_{10}(j_{g}) - log_{10}(M_{g})$ relations. 
The values of $j$ for the stars, gas and the baryons in FGC 1440 are consistent with those of normal spiral galaxies with similar mass.  

\end{itemize}

\section{Data Availability}
The data from this study are available upon request.

\section{ACKNOWLEDGEMENT}
We thank the staff of the GMRT that made these observations possible. GMRT is run by the National Centre for Radio Astrophysics of the Tata Institute of Fundamental Research. 
PK is  partially  supported  by  the  BMBFproject 05A17PC2 for D-MeerKAT. SB, AA, DM acknowledge the support by the Russian Science Foundation, grant 19-12-00145. 
We acknowledge the usage of the HyperLeda database (\textcolor{blue}{http://leda.univ-lyon1.fr}). We thank  Peter Yoachim for kindly providing the optical rotation curve of
FGC 1440.

\small{\bibliographystyle{mnras}}
\bibliography{1440,ref2} 

\section{Appendix-A}
\begin{figure*}
%\hspace{-10mm}
\resizebox{150mm}{49mm}{\includegraphics{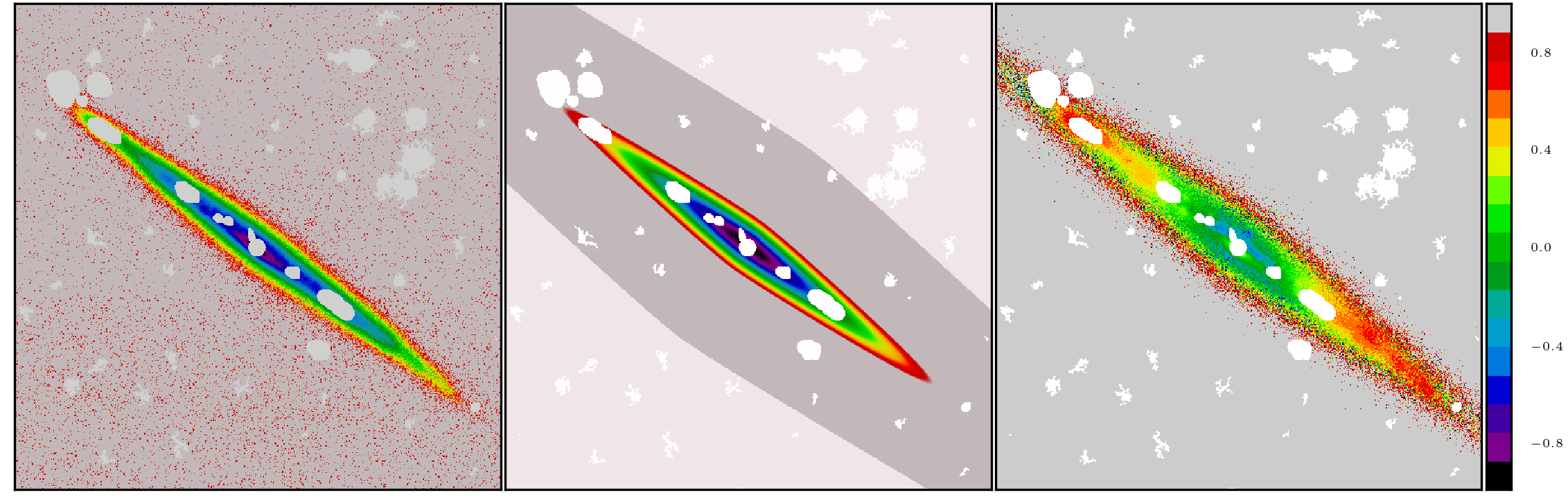}} 
\caption{ The image shows the data (panel 1), the Galfit model (panel 2) and the nomalized difference 
between the data and the model (panel 3) for optical photometery of FGC 1440 using the SDSS g-band image. }
\end{figure*}
\begin{figure*}
%\hspace{-10mm}
\resizebox{150mm}{49mm}{\includegraphics{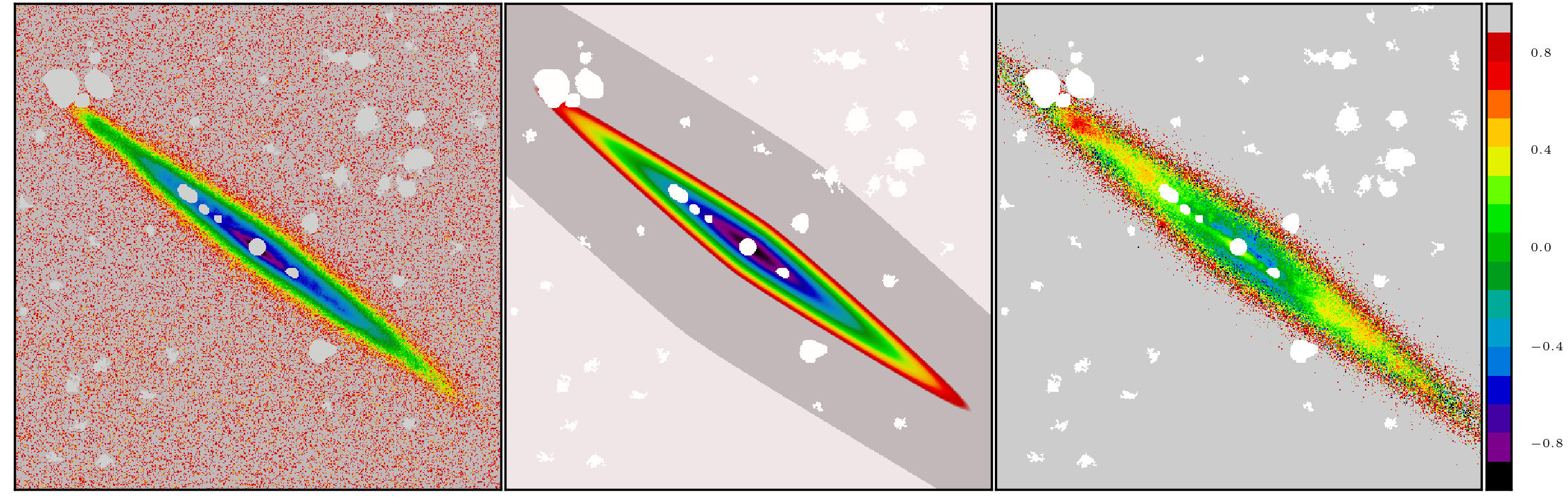}} 
\caption{ The image shows the data (panel 1), the Galfit model (panel 2) and the nomalized difference 
between the data and the model (panel 3) for optical photometery of FGC 1440 using the SDSS r-band image.} 
\end{figure*}
\begin{figure*}
%\hspace{-10mm}
\resizebox{150mm}{49mm}{\includegraphics{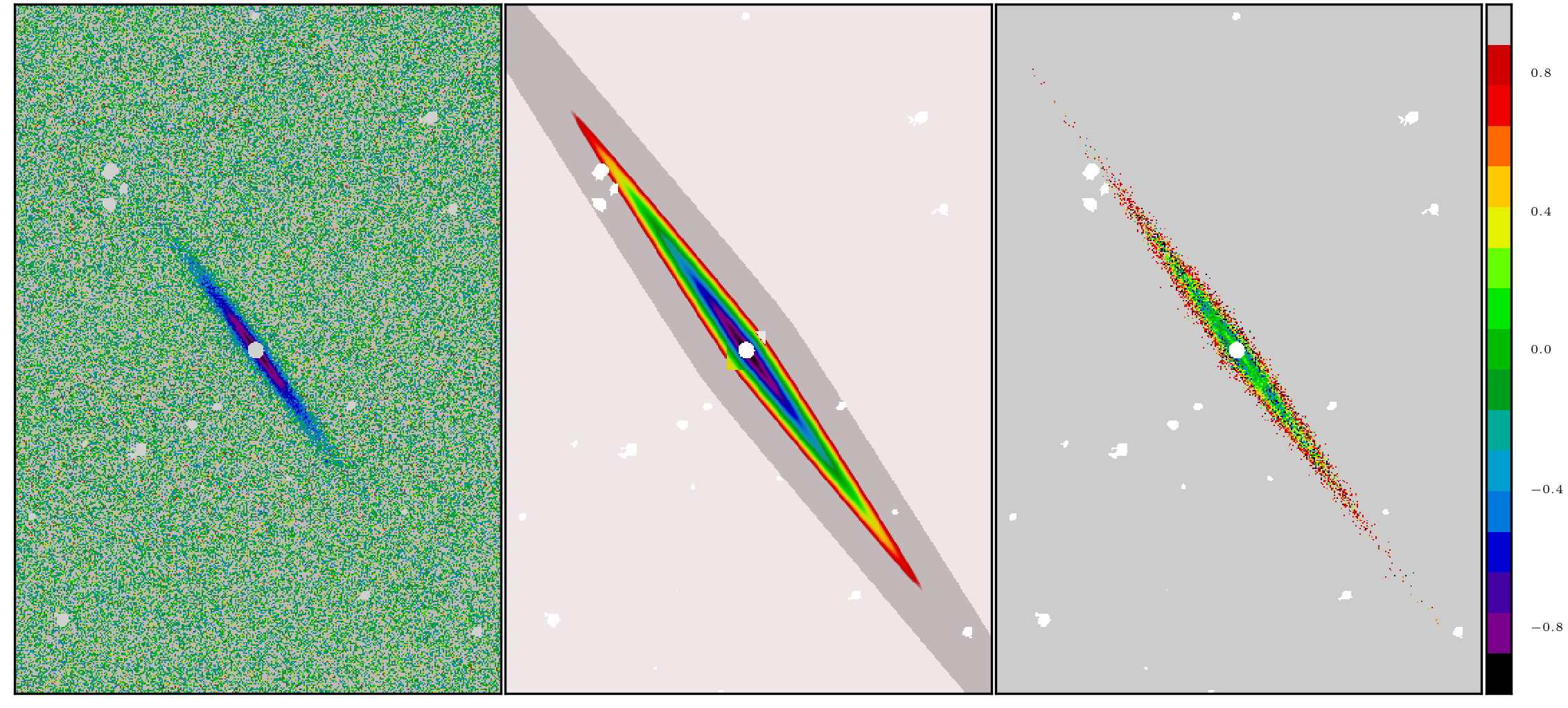}} 
\caption{ The image shows the data (panel 1), the Galfit model (panel 2) and the nomalized difference 
between the data and the model (panel 3) for optical photometery of FGC 1440 using the UKIDSS K-band image.} 
\end{figure*}
\section{Appendix-B}
\begin{figure*}
\resizebox{190mm}{180mm}{\includegraphics{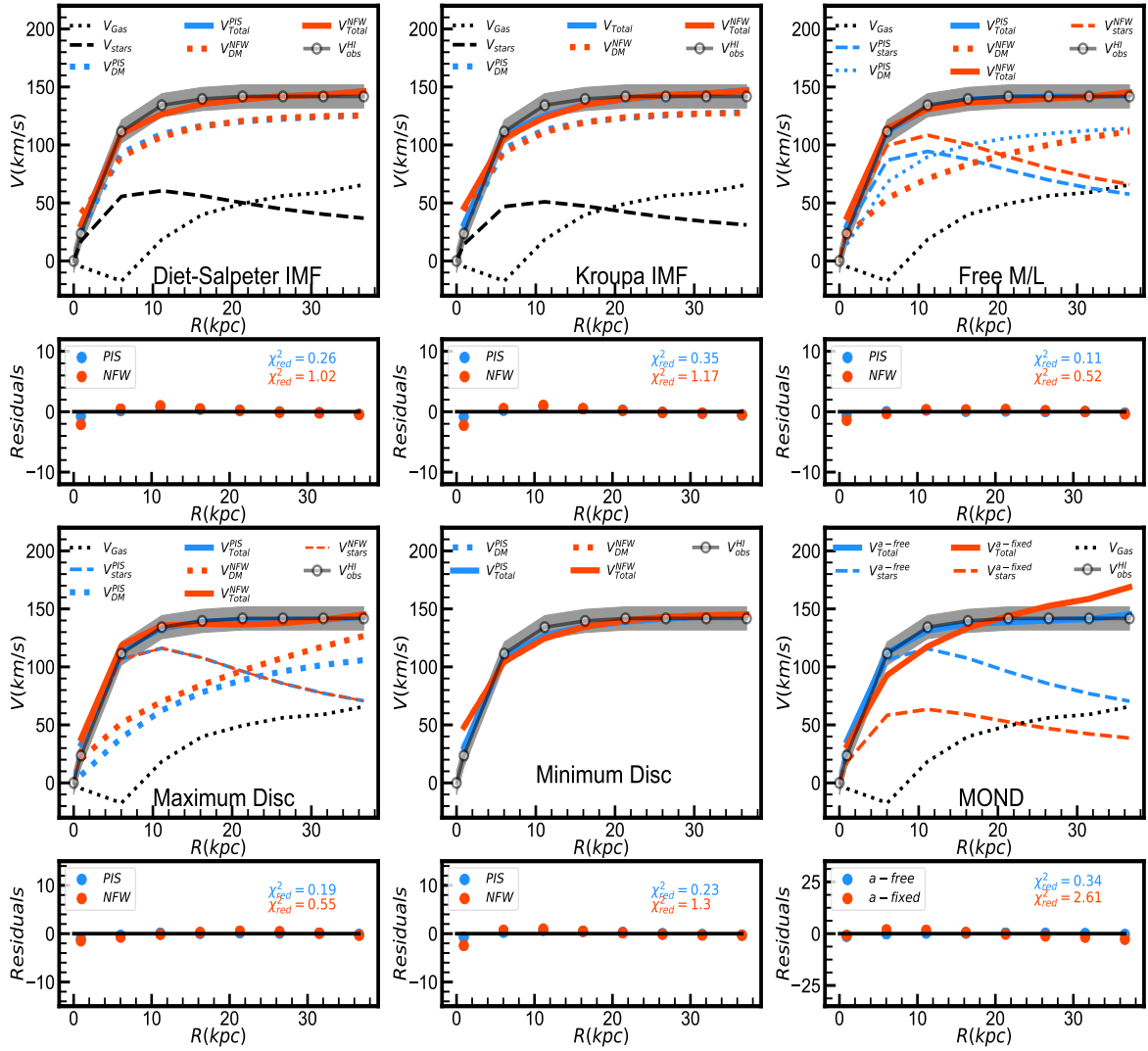}} 
\caption{ We present the mass-model of the galaxy FGC 1440 derived using SDSS g-band photometry. The mass models are constrained using the
\HI{}  21 cm data.}
\end{figure*}
\begin{figure*}
\resizebox{190mm}{180mm}{\includegraphics{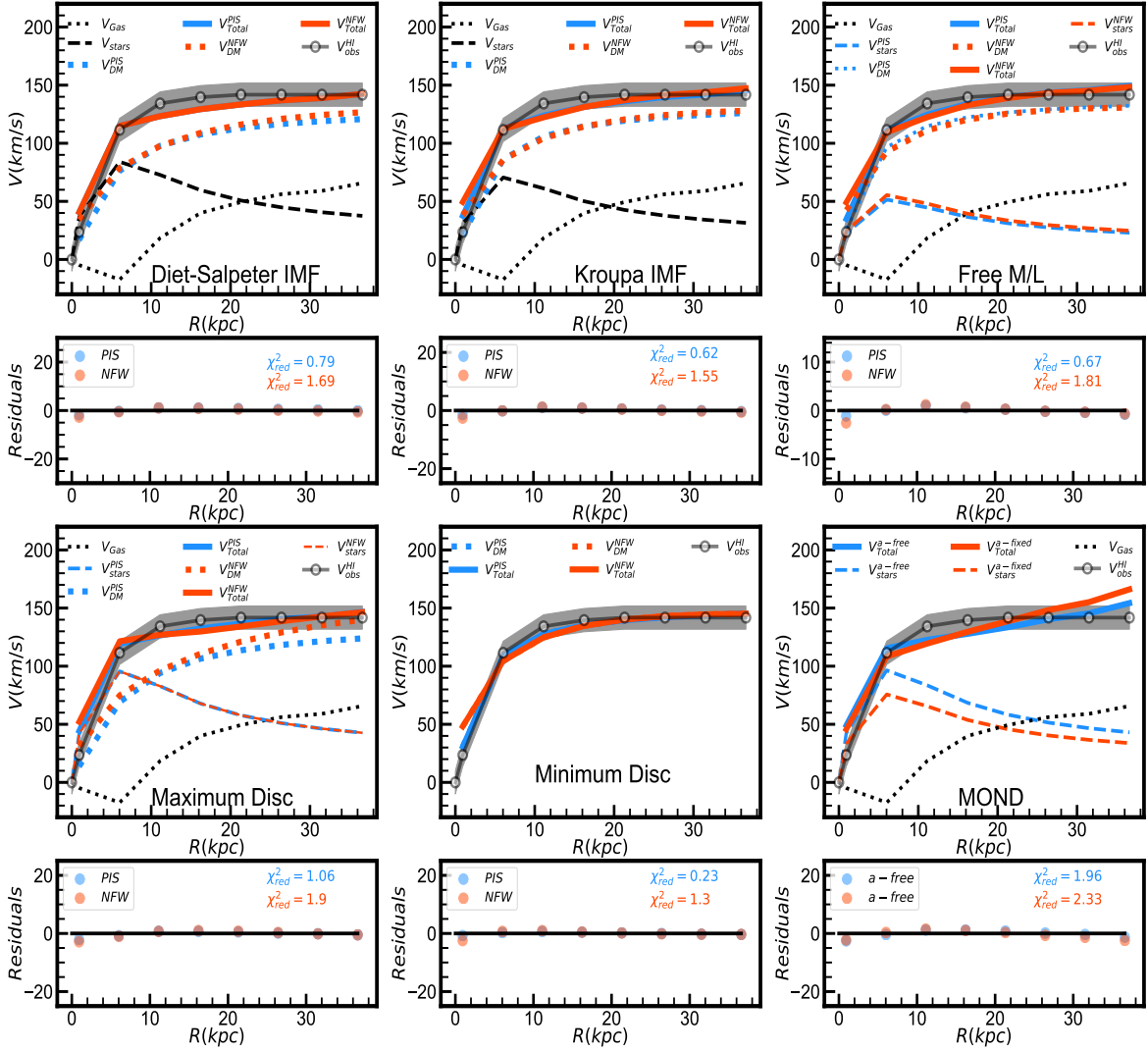}} 
\caption{We present the mass-model of the galaxy FGC 1440 derived using UKIDSS K-band photometry. The mass models are constrained using the
\HI{}  21 cm data.}
\end{figure*}
\begin{table*}
\hspace{-65.5mm}
\begin{minipage}{110mm}
\hfill{}
\caption{Dark matter density parameters derived from mass-modeling using the optical g-band and NIR K band photometry using just the \HI{} rotation curve.}
\centering
\begin{tabular}{|l|c|c|c|c|c|c|c|c|c|c|}
\hline
\hline
Model             & $c^{\textcolor{red}{(a)}}$           &$R^{\textcolor{red}{(b)}}_{200}$         & $\gamma^{*\textcolor{red}{(c)}}$&$\frac{V_{max}}{V_{200}}^{\textcolor{red}{(d)}}$     & $\chi^{2\textcolor{red}{(e)}}_{red}$     & $\rho^{\textcolor{red}{(f)}}_{0}\times10^{-3}$  &$R^{\textcolor{red}{(g)}}_{c}$ & $\gamma^{*\textcolor{red}{(h)}}$& $\frac{R_{c}}{R_{d}}^{\textcolor{red}{(i)}}$     &$\chi^{2\textcolor{red}{(j)}}_{red}$ \\
                  &             &(kpc)             &             &              &                  &  $M_{\odot}/pc^{3}$      &(kpc)   &   \\        &      &                                  \\
g-band            & NFW  profile&                  &             &              &                &PIS profile               &        &             &      & \\
\hline
$'diet'$ Salpeter	&$	4.65\pm1.18	$  &$	87.15\pm7.2$	&$	3.8	$&$	2.23	$&$	1.02	$&$	45.92\pm12.7$&$	2.66\pm0.42	$&$	3.8	$&$	0.60	$&$	0.26$	\\  
Kroupa IMF 	&$	5.10\pm1.27	$&$	87.4\pm7.07	$&$	2.6	$&$	2.22	$&$	1.17	$&$	54.78\pm16.4	$&$	2.48\pm0.42	$&$	2.6	$&$	0.56	$&$	0.35$	         \\
Free $\gamma^{*}$ 	&$	1.06\pm1.43	$&$	103.5\pm35.8$	&$	12.19\pm2.84	$&$	1.87	$&$	0.52	$&$	13.08\pm8.65	$&$	4.76\pm1.60	$&$	9.26\pm2.10	$&$	1.07	$&$	0.11$	\\
Maximum Disc 	&$	0.05\pm0.74	$&$	221.36\pm179.9	$&$	14	$&$	0.87	$&$	0.55	$&$	2.7\pm0.86	$&$	11.19\pm2.7	$&$	14	$&$	2.53	$&$	0.19$	\\
Minimum Disc	&$	5.4\pm1.16	$&$	98.7\pm7.23	$&$	0	$&$	1.97	$&$	1.3	$&$	65.81\pm13.04	$&$	2.56\pm0.29	$&$	0	$&$	0.57	$&$	0.23	$\\
\hline
K-Band            &                    &                      &   & &                      &   &\\
\hline
$'diet'$ Salpeter	&$	3.13\pm1.3	$&$	94.3\pm13.95$	&$	0.85	$&$	2.06	$&$	1.69	$&$	18.69\pm8.63	$&$	4.15\pm1.22	$&$	0.85	$&$	1.60	$&$	0.79$	\\
Kroupa IMF 	&$	3.94\pm1.32	$&$	91.38\pm10.51$	&$	0.6	$&$	2.12	$&$	1.55	$&$	29.035\pm11.81	$&$	3.42\pm0.84	$&$	0.6	$&$	1.3	$&$	0.62$	\\
Free $\gamma^{*}$ 	&$	4.72\pm2.73	$&$	90.71\pm10.37$	&$	0.37\pm0.63	$&$	2.14	$&$	1.81	$&$	46.7\pm34.2	$&$	2.79\pm1.04	$&$	0.32\pm0.36	$&$	1.08	$&$	0.67$	\\
Maximum Disc 	&$	2.10\pm1.14	$&$	117.56\pm26.3	$&$	1..1	$&$	1.65	$&$	1.9	$&$	12.16\pm6.46	$&$	5.43\pm1.83	$&$	1.1	$&$	2.10	$&$	1.06$	\\
Minimum Disc	&$	5.41\pm1.61	$&$	98.7\pm7.23	$&$	0	$&$	1.97	$&$	1.3	$&$	65.81\pm13.04	$&$	2.57\pm0.29	$&$	0	$&$	0.99	$&$	0.23$	\\
\hline

MOND              &                      &              &              &                &             &    &    \\
\hline
& $a^{\textcolor{red}{(k)}}$                    &$\gamma^{*\textcolor{red}{(l)}}$  &  $\chi^{2\textcolor{red}{(m)}}_{red}$  &                &             &    & &  \\
                  & $m^{2}s^{-2}$   &              &              &                &             &    &    \\        &                                        \\
\hline
g-Band              &$0.4 \times 10^{-10}$   &$13.89\pm1.64 $      & 0.34            &                &             &    &              \\    
$g-Band^{a=fixed}$  &$1.2 \times 10^{-10}$   &$ 2.61  $            &              &             &    &               \\
K-Band              &$0.8 \times 10^{-10}$   &$ 1.12\pm0.4$        &1.96&                &             &    &               \\                     
$K-Band^{a=fixed}$  &$1.2 \times 10^{-10}$   &$ 0.69\pm0.18$    & 2.33              &                &             &    &               \\
\hline
\end{tabular}  
\hfill{}
\label{table: table 5}
\end{minipage}
\begin{tablenotes}
\item  $\textcolor{red}{(a)}$:  Concentration parameter of the NFW profile\\
\item $\textcolor{red}{(b)}$:  Radius at which the mean density equalt 200 times the critical density.\\
\item $\textcolor{red}{(c)}$:  Mass to light ratio derived using population synthesis models or estimated as a free parameter. \\
\item $\textcolor{red}{(d)}$:  Ratio of the asymptotic velocity to the  velocity at $R_{200}$, where $\frac{V_{200}}{\kms}=0.73\frac{R_{200}}{kpc}$\citep{navarro1997universal}\\
\item $\textcolor{red}{(e)}$:  Reduced chi-square value corresponding to the fit. \\
\item $\textcolor{red}{(f)}$:  The central dark matter density of the PIS dark matter halo model\\
\item $\textcolor{red}{(g)}$:  The core radius of the PIS dark matter halo model\\
\item $\textcolor{red}{(h)}$:  Mass to light ratio derived using population synthesis models or estimated as a free parameter. \\
\item $\textcolor{red}{(i)}$:  Ratio of the core radius and the disk scalelenght.\\
\item $\textcolor{red}{(j)}$:  Reduced chi square value corresponding to the fit.\\
\item $\textcolor{red}{(k)}$:  Acceleration per lenght in MOND.\\
\item $\textcolor{red}{(l)}$:  Estimated Mass to light ratio in MOND.\\
\item $\textcolor{red}{(m)}$:  Reduced chi square corresponding to the fit.\\
\end{tablenotes}
\end{table*}

\end{document}